\documentclass[fleqn,usenatbib]{mnras}

\usepackage{newtxtext,newtxmath}

\usepackage[T1]{fontenc}
\usepackage{ae,aecompl}
\usepackage{siunitx}
\usepackage{soul}

\usepackage{graphicx}	
\usepackage{amsmath}	
\usepackage{amssymb}	
\usepackage{pdflscape}
\usepackage{booktabs}
\usepackage{threeparttable}
\usepackage{hhline}

\usepackage{varwidth}

\usepackage{datetime}

\begin{document}
\label{firstpage}
\pagerange{\pageref{firstpage}--\pageref{lastpage}}
\title[MOA-II Eclipsing Binary Catalogue]{The First Eclipsing Binary Catalogue from the MOA-II database}

\author[M.~C.~A.~Li et al.]
{\parbox{\textwidth}{M.~C.~A.~Li,$^{1}$\thanks{E-mail: \texttt{mli351@aucklanduni.ac.nz}}
N.~J.~Rattenbury,$^{1}$
I.~A.~Bond,$^{2}$
T.~Sumi,$^{3}$
D.~P.~Bennett,$^{4}$
N.~Koshimoto,$^{3}$
F.~Abe,$^{5}$
Y.~Asakura,$^{5}$
R.~Barry,$^{6}$
A.~Bhattacharya,$^{4}$
M.~Donachie,$^{1}$
P.~Evans,$^{1}$
M.~Freeman,$^{7}$
A.~Fukui,$^{8}$
Y.~Hirao,$^{3}$
Y.~Itow,$^{5}$
C.~H.~Ling,$^{2}$
K.~Masuda,$^{5}$
Y.~Matsubara,$^{5}$
Y.~Muraki,$^{5}$
M.~Nagakane,$^{3}$
K.~Ohnishi,$^{9}$
To.~Saito,$^{10}$
A.~Sharan,$^{1}$
D.~J.~Sullivan,$^{11}$
D.~Suzuki,$^{4}$
P.~J.~Tristram$^{12}$ and
A.~Yonehara$^{13}$}\vspace{0.4cm}\\
\parbox{\textwidth}{$^{1}$Department of Physics, University of Auckland, Private Bag 92019, Auckland, New Zealand\\
$^{2}$Institute of Natural and Mathematical Sciences, Massey University, Private Bag 102-904, North Shore Mail Centre, Auckland, New Zealand\\
$^{3}$Department of Earth and Space Science, Graduate School of Science, Osaka University, 1-1 Machikaneyama, Toyonaka, Osaka 560-0043, Japan\\
$^{4}$Department of Physics, University of Notre Dame, Notre Dame, IN 46556, USA\\
$^{5}$Solar-Terrestrial Environment Laboratory, Nagoya University, Nagoya, 464-8601, Japan\\
$^{6}$Astrophysics Science Division, NASA Goddard Space Flight Center, Greenbelt, MD 20771, USA\\
$^{7}$School of Physics, The University of New South Wales, Sydney NSW 2052, Australia\\
$^{8}$Okayama Astrophysical Observatory, National Astronomical Observatory, 3037-5 Honjo, Kamogata, Asakuchi, Okayama 719-0232, Japan\\
$^{9}$Nagano National College of Technology, Nagano 381-8550, Japan\\
$^{10}$Tokyo Metropolitan College of Industrial Technology, Tokyo 116-8523, Japan\\
$^{11}$School of Chemical and Physical Sciences, Victoria University, Wellington, New Zealand\\
$^{12}$University of Canterbury, Mount John Observatory, P.O. Box 56, Lake Tekapo 8770, New Zealand\\
$^{13}$Department of Physics, Faculty of Science, Kyoto Sangyo University, 603-8555 Kyoto, Japan}}

\date{Accepted XXX. Received YYY; in original form ZZZ}

\pubyear{2017}

\maketitle

\begin{abstract}
We present the first catalogue of eclipsing binaries in two MOA fields towards the Galactic bulge, in which over 8,000 candidates, mostly contact and semi-detached binaries of periods $<1\,\text{d}$, were identified. In this paper, the light curves of a small number of interesting candidates including eccentric binaries, binaries with noteworthy phase modulations and eclipsing RS CVn type stars are shown as examples. In addition, we identified three triple object candidates by detecting the light-travel-time effect in their eclipse time variation curves.
\end{abstract}




\section{Introduction}

The present evolution of the Microlensing Observations in Astrophysics (MOA-II) project began in 2006. The MOA-II survey observes towards the Galactic bulge (GB) to detect microlensing signals which show additional perturbations associated with a planetary object in the lens star system \citep{2013ApJ...778..150S}. Using a 1.8m telescope with a wide-field CCD camera, MOA-II monitors millions of stars simultaneously towards the densely populated regions of the GB. The rewards of collaborative efforts by MOA-II and other microlensing research groups include the discovery of over 50 exoplanets \citep{2016ApJ...833..145S} and evidence for free-floating planets, which are not orbiting around any star \citep{2016A&A...595A..53B}. The MOA-II project has also resulted in a large amount of photometric data, spanning over 8 years to date for  variable objects which are worth categorization for future study.

Eclipsing binaries (EBs) are particularly interesting and useful variable objects. Stellar binaries are common in our Galaxy and they might be the origins of many astrophysical phenomena such as supernova explosions, gamma-ray bursts and accretion disks, etc. \citep{2017MNRAS.465.4650K, 2016SSRv..202...33L, 2017arXiv170206160C, 2017PASA...34....1D}. However determining the properties of a stellar binary, even identifying a binary system, is usually unachievable without an observed periodicity either in photometry or spectroscopy. In this sense, EBs are crucial as we can determine their orbital periods by observing the eclipses repeatedly. Combining this with high-quality radial velocity measurements using spectroscopy can further allow a complete modelling of an EB's light curve, and a determination of fundamental parameters, such as temperature, mass, radius and luminosity of each stellar component with high accuracy, which can be used to test current stellar models \citep{2016AJ....151..139M}.

EBs can be used to detect orbiting companions by measuring variations in eclipse times via the light-travel-time effect (LTTE) \citep{1990BAICz..41..231M}. Over 200  EBs discovered by the \textit{Kepler} space telescope were identified as triple systems via the eclipse time variation (ETV) method \citep{2016MNRAS.455.4136B}. The ETV method was also successfully applied to search for circumbinary planets in post-common-envelope binary (PCEB) systems (e.g. \citealt{2015A&A...577A.146B}, \citealt{2014MNRAS.445.2331L}). The measurement of ETV, as pointed out in a few papers on circumbinary planets (e.g. \citealt{2016MNRAS.455L..46M}, \citealt{2016ApJ...831...96L}), is a plausible method to unveil any circumbinary planet with a highly inclined orbit with respect to the orbital plane of its host binary, which would be hidden from the transit and radial velocity methods.

In addition, long-term observations of EBs can provide an opportunity to detect the change in the orbital state and probe the interior of an EB; for example, the detection of apsidal motion in an eccentric binary by accurate eclipse timings can act as an indirect way to determine the internal structure of the star, allowing stellar and evolution model testing (see, e.g., \citealt{1992A&AS...96..255C} and references therein). Since every essential orbital and stellar parameter of EBs can be potentially determined from photometry and spectroscopy, their distances can be potentially derived as well. EBs can, therefore, serve as distance estimators of local galaxies (e.g. \citealt{2014ApJ...780...59G}, \citealt{2013Natur.495...76P} and \citealt{2010A&A...509A..70V}) and stellar clusters (e.g. \citealt{2009AJ....137.5086M}), and be used to study the spatial distribution of stars in the Galaxy \citep{2013MNRAS.432.2895H}.

Given the interesting and special role of EBs amongst different types of variable stars, most wide-field survey projects have made efforts to identify and catalogue EBs in their databases. And, over the last decade, the number of EBs identified in the Galaxy has been dramatically increased thanks to ground-based and space wide-field surveys, including the Optical Gravitational Lensing Experiment (OGLE) and the NASA \textit{Kepler} mission. The latter published the \textit{Kepler} Eclipsing Binary Catalog that contains over 2,800 EBs in the \textit{Kepler} fields \citep{2016AJ....151...68K}, while the OGLE EB catalogue contains over 450,000 and 11,589 EBs towards the GB \citep{2016AcA....66..405S} and Galactic disk fields \citep{2013AcA....63..115P}, respectively. The triumph of such works, in addition to a sheer number of various types of new EBs, is the discovery of a new class of eccentric binaries, called the heartbeat stars, which show periodic pulsations arising from the tidal interaction between two stellar components \citep{2012ApJ...753...86T}. 

MOA-II has dedicated most of its telescope time to observing 22 fields towards the GB, and collected $\sim$100 terabytes ($\text{TBs}$) of image data to date. We present in this paper the first EB catalogue, containing over 8,000 EBs from two MOA fields. In Section~\ref{sec:obs_dat} we provide the details of the observations and the treatment of data. In Section~\ref{sec:iden_eb} we describe our strategy to identify EB candidates. The examples of interesting EB candidates in our catalogue and the results in the ETV analysis of three selected EBs are presented in Section~\ref{sec:results}. We discuss and conclude our work in Section~\ref{sec:discuss}.

\begin{figure}
	\includegraphics[width=\columnwidth]{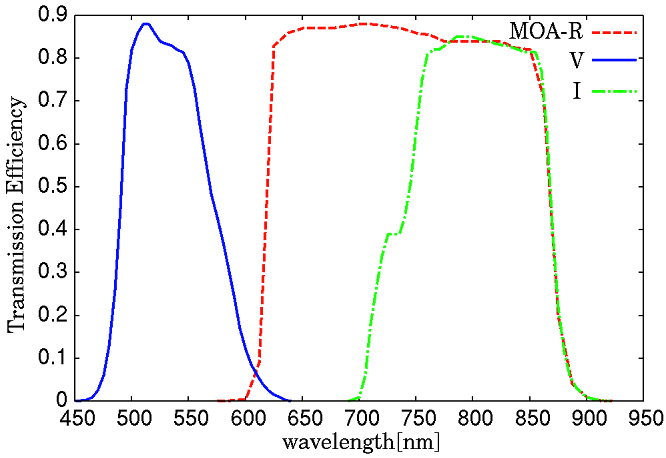}
    \caption{The MOA passbands. The MOA-R band spanning from 600$\,$nm to 900$\,$nm is routinely used for the MOA-II observations. As the microlensing effect is independent of wavelength, the V-band observations are only carried out occasionally in order to obtain the colour difference information of microlensing source stars. In addition, as the MOA-R band already covers the I band, the I band is no longer used.}
    \label{fig:moa_filter}
\end{figure}

\section{Observations and Data Reductions}
\label{sec:obs_dat}
The MOA-II observations towards the GB fields have been routinely carried out from the end of February to the beginning of November every year since 2006, using the 1.8m MOA-II telescope located at the University of Canterbury Mount John Observatory, New Zealand. The wide-field camera, MOA-cam3, consisting of ten \(2\,\text{k}\,\times\,4\,\text{k}\) pixel CCDs with \(15\,\mu\)m pixels, is installed on the telescope, providing a pixel scale of 0$_{.}^{''}$58 pixel$^{-1}$ that gives a field of view (FOV) of 2.18 deg$^{2}$ \citep{2008ExA....22...51S}. The MOA-II telescope takes the images of 22 fields towards the GB with high cadences, through the custom MOA-Red wide-band filter, which is roughly equivalent to the combination of the standard Kron/Cousins R and I bands, from \(600\,\)nm to \(900\,\)nm (see Figure~\ref{fig:moa_filter}). The exposure time of a MOA image of a GB field is $60\,$s. 

The difference imaging analysis (DIA) method was used to reduce the MOA images. The DIA method extracts variable objects in an observed image by subtracting the observed image from a high quality, good seeing, reference image that is transformed to have the same seeing and photometric scaling as the observed image beforehand (for details see the supplement in \citealt{2013ApJ...778..150S}, and \citealt{2001MNRAS.327..868B}). As a result, the DIA method gives the measurement of the relative fluxes instead of the absolute fluxes. It makes the DIA method intrinsically good at detecting transient events and faint variable objects. The DIA method achieves photometry with better precision than other methods towards very crowded fields, e.g. the GB fields, because almost all stars in such fields cannot be resolved individually, in practice, by straight PSF-fitting routines such as DOPHOT36. It is verified by re-analyzing the MACHO and OGLE databases that the detection rates of microlensing events are significantly improved by the DIA method which overcomes the problems of undetectability due to the blending and faintness of the microlensing source stars in the crowded fields \citep{1999ApJS..124..171A, 1999A&A...343...10A}.

\begin{table}
	\centering
	\caption{The coordinates of the centers of the MOA Galactic bulge (GB) fields. The first column shows the MOA GB field numbers, the fourth column gives the number of images,  $N_{\text{image}}$, for a GB field taken from February 2013 to August 2014, and $t_{\text{exp}}$ is the exposure time.}
	\label{tab:coord_moa_fields}
	\begin{tabular}{lcccc} 
		\hline
		GB & R.A.$\,$(J2000) & Dec$\,$(J2000) & $N_{\text{image}}$ & $t_{\text{exp}}$\\
        & (h:m:s) & (d:m:s) & & (sec)\\
		\hline
        $\ $1 & 17:47:31.4 & -34:14:31.1 & 2559 & 60\\
        $\ $2 & 17:54:01.4 & -34:29:31.1 & 2415 & 60\\
        $\ $3 & 17:54:01.4 & -32:44:31.1 & 6579 & 60\\
        $\ $4 & 17:54:01.4 & -30:59:31.1 & 6456 & 60\\
        $\ $5 & 17:54:01.4 & -29:14:31.1 & 6625 & 60\\
        $\ $6 & 17:54:01.4 & -27:29:31.1 & 443 & 60\\
        $\ $7 & 18:00:01.4 & -32:44:31.1 & 1324 & 60\\
        $\ $8 & 18:00:01.4 & -30:59:31.1 & 2439 & 60\\
        $\ $9 & 18:00:01.4 & -29:14:31.1 & 6310 & 60\\
        $\ $10 & 18:00:01.4 & -27:29:31.1 & 6085 & 60\\
		$\ $11 & 18:06:01.4 & -32:44:31.1 & 1051 & 60\\
        $\ $12 & 18:06:01.4 & -30:59:31.1 & 1205 & 60\\
        $\ $13 & 18:06:01.4 & -29:14:31.1 & 2345 & 60\\ 
        $\ $14 & 18:06:01.4 & -27:29:31.1 & 5952 & 60\\
        $\ $15 & 18:06:01.4 & -25:44:31.1 & 1093 & 60\\
        $\ $16 & 18:12:01.4 & -29:14:31.1 & 1179 & 60\\
        $\ $17 & 18:12:01.4 & -27:29:31.1 & 2092 & 60\\
        $\ $18 & 18:12:01.4 & -25:44:31.1 & 2034 & 60\\
        $\ $19 & 18:18:01.4 & -25:29:31.1 & 981 & 60\\
        $\ $20 & 18:18:01.4 & -23:44:31.1 & 1167 & 60\\
        $\ $21 & 18:18:01.4 & -21:59:31.1 & $--$ & 60\\
        $\ $22 & 18:36:25.4 & -23:53:31.1 & $--$ & 60\\
		\hline
	\end{tabular}
\end{table}

MOA-II has collected $\sim$100 $\text{TBs}$ of image data of the GB fields since 2006. In order to reduce the pressure on data storage, the light curves, i.e. the time series of photometric measurements, of all resolved stars are not maintained in the database, but their positions in the subtracted images are. The light curves of variable objects thus will only be produced if requested. And, when requested, their photometric measurements, i.e. the relative fluxes, are extracted from the subtracted images using aperture photometry with an aperture radius of 6 pixels. Generating the light curves of every star in a GB field from the full MOA-II database is time-consuming and for this reason, and for the purposes of demonstrating our method, only fields GB9 and GB10 were inspected to search for EB candidates using the data of two MOA observational seasons, collected from February 2013 to August 2014. The coordinates of the centers of all GB fields are listed in Table~\ref{tab:coord_moa_fields}. The GB9 and GB10 fields were selected for inspection simply because these fields were observed most frequently. They overlap with the OGLE fields \citep{2011AcA....61...83S} which allows us to cross reference the two databases.

\section{Identification of Eclipsing Binaries}
\label{sec:iden_eb}
The algorithm used by MOA-II to extract variable stars from the subtracted images yielded $\sim$8,000,000 variable objects in the two GB fields we inspected. Machine learning or semi-automatic algorithms are desirable to search for EB candidates in such a great deal of data. Unfortunately, the datasets also included spurious variable objects which result from contamination by nearby variable stars, or imperfect image subtraction at the stars' positions due to the effects of differential refraction \citep{2001MNRAS.327..868B,1999ApJS..124..171A}. This made the existing automatic classification algorithms used by other research groups such as OGLE (e.g. \citealt{2006MNRAS.367.1531M}) and \textit{Kepler} (e.g. \citealt{2016MNRAS.456.2260A}) teams less applicable to our datasets. We did not attempt therefore to develop our own automatic or semi-automatic algorithm to do the task, given that no EB catalogue of MOA-II existed before for testing. Instead, we adopted the straightforward and tedious strategy; that is, we performed a period analysis of the datasets and folded the light curves at the calculated periods, and we then identified EB candidates by inspecting their shapes by eye.

\subsection{Cleaning Light Curves}
Before finding periodic signals, we first cleaned every light curve by iterating twice an outlier rejection algorithm that filtered out the data points with values \(4.0\sigma\) above or \(9.0\sigma\) below the relative flux mean, and detrended the light curves by linear regression. The asymmetric criterion in relative flux was taken to avoid rejecting data points corresponding to an eclipse. As we expected the error bar of a measurement point should not be much larger than the overall relative flux standard deviation, data points with errors \(>\sqrt{3}\sigma\) were discarded as unreliable measurements. Note that $\sigma$ is the standard deviation of all relative flux values in the light curve. After cleaning, and checking a certain number of the resultant light curves, we ignored the light curves with $<1000$ good data points as being poor objects.

\subsection{Period Analysis} \label{sec:pa}
\subsubsection{Conditional Entropy Method}
As we wanted to take all good data points of a light curve into account to search for periodic signals, we needed a period finding algorithm (PFA) that is still relatively fast, and of course robust, even when handling large time series data. The conventional PFAs such as the Lomb-Scargle periodogram (\citealt{1976Ap&SS..39..447L} and \citealt{1982ApJ...263..835S}), Fourier transform \citep{1981AJ.....86..619F} and the phase dispersion minimization (PDM) method \citep{1978ApJ...224..953S} are not satisfactory in terms of their computational speeds with a large amount of input data such as our datasets which contain over 3,000 measurement points per light curve. We thus adopted the conditional entropy (CE) method proposed by \cite{2013MNRAS.434.2629G} to determine the light curve periods.

The CE method is a modified version of the Shannon entropy method that was first introduced by \citet{1995ApJ...449..231C}. It determines the period of a light curve by minimizing the conditional entropy over a range of trial periods at which the light curve is folded, based on the idea that the most ordered arrangement of a light curve, corresponding to the signal with the highest recognizable shape, in the phase-flux plane, should be established when it is folded at the correct period. The conditional entropy, \(H_{c}=H(m|\phi)\), which should be minimized is defined by:
\begin{equation}
    H_{c}=\sum_{i,j}p(m_{i},\phi_{j})\ln\bigg(\frac{p(\phi_{j})}{p(m_{i},\phi_{j})}\bigg),
	\label{eq:ce}
\end{equation}
where \(p(m_{i},\phi_{j})\) is the occupation probability for the \(i^{th}\) partition in flux and the \(j^{th}\) partition in phase and \(p(\phi_{j})\) is the occupation probability of the \(j^{th}\) phase partition, which for rectangular partitioning is just
\begin{equation}
    p(\phi_{j})=\sum_{i}p(m_{i},\phi_{j}).
	\label{eq:prob_phi_j}
\end{equation}
We divided the phase-flux plane into $20\times20$ grids of equal size. The boundaries of the plane were taken as the maximum and minimum fluxes of the light curve. The trial periods ranged from $0.05$ day to 600 days. The occupation probability was simply calculated by counting the number of points in the grids involved over the total number of points. The CE method gave faster performance than the PDM method with a large number of data points, while they both produced similar results for almost every light curve in the unpublished catalogue of unclassified MOA variable objects.

\subsubsection{Phase Dispersion Minimization Method}
However, the conditional entropy is not a useful measure to rank the periodic signals of different light curves in order to easily pick out obvious EB candidates, because the minimized conditional entropy values just tell us how compact the folded light curves are but do not give an absolute sense regarding the sharpness of their shapes. As a consequence, the PDM statistic, $\Theta$, defined by:
\begin{equation}
	\Theta = s^{2}/\sigma^{2},
    \label{eq:pdm}
\end{equation}
was calculated for the light curve folded at the period determined by the CE method as the ranking measure, where \(s^{2}\) is the overall variance of the partitions of the folded light curve in phase, given by
\begin{equation}
	s^{2} = \frac{\sum_{j=1}^{M}(n_{j}-1)s_{j}^{2}}{\sum_{j=1}^{M}{n_{j}} - M},
    \label{eq:s2}
\end{equation}
and \(\sigma^{2} = \sum_{i=1}^{N}(x_{i} - \bar{x})^{2}/(N-1)\) is the variance of the folded light curve of $N$ data points. Note that, in eq.(\ref{eq:s2}), \(s_{j}^{2}\) and \(n_{j}\) are the variance and number of points in the \(j^{th}\) partition in phase, respectively, and \(M\) is the total number of partitions, i.e. the number of bins in phase. We used \(M=20\) bins in our calculations. The PDM statistic, $\Theta$, is appropriate for our purpose since it intrinsically measures how much the shape of a folded light curve is different from a horizontal straight line. The larger the value of $\Theta$, the more identical the folded light curve is to a horizontal straight line. Therefore, those folded light curves with the lowest values of $\Theta$ would go to the top in our list of candidates. One thing that should be mentioned is that the observation time of a MOA image was taken as the midpoint between the start and end times of exposure, which were recorded up to 6 decimal places in Julian Date. The difference between Julian Date (JD) and Barycentric Julian Date (BJD), i.e. $\text{BJD} - \text{JD} = \Delta_{R\sun} + \Delta_{C}$, where $\Delta_{R\sun}$ is the R{\o}mer delay (i.e. the time correction when the barycentre of the Solar system is taken as the reference point) and $\Delta_{C}$ is the clock correction from Coordinated Universal Time (UTC) to Barycentric Dynamical Time (TDB\footnote{TDB is the abbreviation for the French term ``\,Temps Dynamique Barycentrique\,". The clock correction from UTC to TDB is given by $\Delta_{C}=N+32.184+\delta$, where $N$ is the number of leap seconds and $\delta$ is the correction from Terrestrial Time (TT) to TDB. The constant term of 32.184 seconds is the correction from International Atomic Time (TAI) to UTC.}), varies over time and can be as much as about 9 minutes to date \citep{2010PASP..122..935E}. Nevertheless, the systematic time errors that are induced when using JD instead of BJD are negligible comparing to a typical period of contact binaries, let alone detached binaries, even though such errors will induce spurious ETVs corresponding to the Earth's heliocentric motion which would smear out any signal associated with the LTTE in an EB. We, therefore, did not convert JD to BJD in the period analysis at the stage of searching for EB candidates. However, we did apply BJD corrections and recalculated the periods of all EB candidates. The eclipsing periods provided throughout this paper are the recalculated periods after the BJD corrections.

\subsection{Light Curve Inspection}

Since the MOA observations, as every ground-based optical observation, always suffer from discontinuity due to day-night cycles which induce \(0.5\,\text{d}\) aliases as the strongest periodic signals in a large majority of our datasets, a way to eliminate the folded light curves of \(0.5\,\text{d}\) aliases was needed. To do this the light curves were folded at double their calculated periods for inspection. And those with phase gaps larger than 0.03, after folding, were ignored. This strategy is workable and helpful first because there were usually sufficient data points covering the phases of our light curves well, even when folded at double the calculated periods, and second because the true periods were often double the calculated periods for EB candidates.

All EB-like folded light curves were listed in the first round of inspection. We then rejected those false positives, which are in fact Cepheids or artifacts, by examining the shapes of the listed light curves more carefully, and determined the multiples of the calculated periods that correspond to the true periods of EB candidates. However, we recognized that many neighbouring EB candidates in our list are exactly or nearly identical in terms of their periods and folded light curve shapes. It indicated that most of the EB candidates we included in the list are spurious. Without reanalyzing the photometry of the light curves, whenever suspicious identical neighbouring EB candidates were found within $6\arcsec$ from each other, we simply chose the one, among them, with the lowest value of PDM statistic, \(\Theta\), as the genuine one.
\label{sec:inspect} 

\begin{figure}
	\includegraphics[width=\columnwidth]{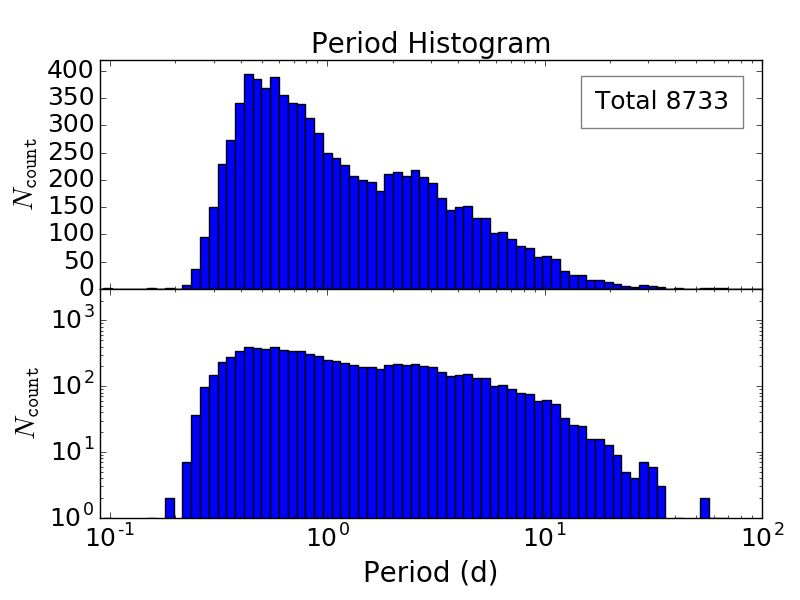}
    \caption{The period histogram of eclipsing binary (EB) candidates in the GB9 and GB10 fields with logarithmic bins. The number of counts, $N_{\text{count}}$, is presented in the linear (upper panel) and logarithmic scale (lower panel), respectively. A total of 8,733 EB candidates were identified in these two fields, using the data from two MOA observational seasons. Among them, 486 EB candidates have no OGLE EB counterpart within $4\arcsec$.}
    \label{fig:period_hist}
\end{figure}

\begin{table*}
	\centering
	\caption{Interesting eclipsing binary (EB) candidates, including eccentric binaries, eclipsing RS Canum Venaticorum (RS CVn) type stars, binaries with noteworthy phase modulations as well as binaries with post main-sequence stellar components, in the GB9 and GB10 fields. The first column shows the MOA number of each EB candidate, the second column shows their corresponding GB fields and the third column gives the CCD chip numbers at which the EB candidates are located. Note that $H_{c}$ is the conditional entropy in eq.(\ref{eq:ce}) and $\Theta$ is the PDM statistic in eq.(\ref{eq:pdm}); $T_{p}$ and $T_{s}$ are the reference epochs of primary and secondary eclipse minima, respectively, measured by the template method in Section~\ref{sec:measure_t_ecl}. The last column shows the OGLE EB counterparts of the MOA EB candidates within 4 arcsec.}
	\begin{tabular}{cccccccccc} 
		\hline
        No. & GB & chip & Period & $H_{c}$ & $\Theta$ & RA & Dec & $|\,{T_{p}-T_{s}}|$ & OGLE Object\\
         & & &(d) & & &(h:m:s)&(d:m:s) &(BJD)& ($<4$ arcsec)\\
        \hline
260591 & 9 & 2 & 1.2581 & 1.3925 & 0.1270 & +17:58:56.82 & -28:55:19.50 & $--$ & $--$\\
263861 & 10 & 4 & 1.7990 & 1.1014 & 0.0586 & +17:58:52.60 & -27:57:16.63 & $--$ & OGLE-BLG-ECL-225277*\\
255102 & 9 & 1 & 2.9331 & 1.2296 & 0.1573 & +17:59:11.29 & -28:32:49.91 & $--$ & OGLE-BLG-ECL-229032**\\
100092 & 10 & 1 & 0.8526 & 1.3759 & 0.5736 & +17:58:45.89 & -26:53:12.64 & $--$ & $--$\\
315721 & 9 & 2 & 3.2436 & 1.2507 & 0.2843 & +17:59:19.52 & -29:01:13.96 & $--$ & OGLE-BLG-ECL-230424*\\
393777 & 9 & 3 & 4.8285 & 1.1393 & 0.5220 & +17:59:37.52 & -29:13:20.04 & $--$ & OGLE-BLG-ECL-233746*\\
23676 & 9 & 7 & 2.1793 & 1.4319 & 0.8533 & +18:02:57.08 & -29:35:22.17 & $--$ & OGLE-BLG-ECL-270404*\\
42742 & 9 & 8 & 2.2359 & 1.3244 & 0.6708 & +18:02:47.89 & -29:19:45.30 & $--$ & OGLE-BLG-ECL-268796*\\
350526 & 9 & 8 & 5.2286 & 1.3003 & 0.0765 & +18:00:39.56 & -29:14:39.34 & $--$ & OGLE-BLG-ECL-245419*\\
349294 & 10 & 7 & 2.3396 & 1.8205 & 0.3710 & +18:00:43.97 & -27:47:42.92 & $--$ & OGLE-BLG-ECL-246259*\\
146114 & 9 & 7 & 3.6452 & 1.1635 & 0.3001 & +18:01:59.68 & -29:26:49.62 & $--$ & OGLE-BLG-ECL-260202*\\
176063 & 10 & 8 & 4.5398 & 1.5166 & 0.3023 & +18:01:49.94 & -27:38:01.43 & $--$ & OGLE-BLG-ECL-258445*\\
57615 & 10 & 1 & 18.2139 & 1.4984 & 0.0822 & +17:58:13.26 & -26:49:45.86 & $--$ & OGLE-BLG-ECL-217710*\\
68403 & 9 & 3 & 19.5865 & 1.0055 & 0.2275 & +17:57:27.67 & -29:22:26.36 & $--$ & OGLE-BLG-ECL-208927*\\
170700 & 9 & 2 & 34.1678 & 1.2526 & 0.0746 & +17:58:18.76 & -28:49:18.71 & $--$ & OGLE-BLG-ECL-218797*\\
312954 & 9 & 2 & 5.5753 & 0.9452 & 0.9098 & +17:59:18.56 & -28:44:26.82 & $--$ & OGLE-BLG-ECL-230294**\\
157137 & 9 & 10 & 6.6259 & 1.3546 & 0.7577 & +18:01:50.26 & -28:25:51.79 & $--$ & OGLE-BLG-ECL-258519*\\
238852 & 9 & 5 & 20.4211 & 1.3200 & 0.5753 & +17:58:57.69 & -29:59:39.40 & $--$ & $--$\\
164457 & 9 & 4 & 3.8996 & 1.1218 & 0.1631 & +17:58:10.09 & -29:31:09.27 & $--$ & OGLE-BLG-ECL-217085*\\
189457 & 9 & 3 & 7.6329 & 1.2755 & 0.2396 & +17:58:17.67 & -29:19:37.28 & $--$ & OGLE-BLG-ECL-218587*\\
282091 & 9 & 9 & 17.6026 & 1.5346 & 0.2778 & +18:00:56.10 & -28:52:33.01 & $--$ & OGLE-BLG-ECL-248557*\\
204923 & 9 & 3 & 5.4672 & 1.9721 & 0.6582 & +17:58:23.83 & -29:23:14.60 & $--$ & OGLE-BLG-ECL-219864*\\
129368 & 10 & 1 & 4.5422 & 1.5755 & 0.6452 & +17:59:05.39 & -26:52:13.39 & $--$ & $--$\\
45015 & 10 & 10 & 6.7270 & 1.3525 & 0.7198 & +18:02:39.23 & -26:51:56.80 & $--$ & OGLE-BLG-ECL-267353*\\
111131 & 9 & 8 & 8.1225 & 1.7073 & 0.8977 & +18:02:17.51 & -29:05:18.55 & $--$ & OGLE-BLG-ECL-263293*\\
147629 & 10 & 1 & 9.0413 & 1.1516 & 0.6541 & +17:59:17.86 & -26:51:13.97 & $--$ & OGLE-BLG-ECL-230190*\\
138632 & 9 & 7 & 19.1310 & 1.3043 & 0.9214 & +18:02:03.19 & -29:40:28.47 & $--$ & OGLE-BLG-ECL-260748*\\
244397 & 9 & 8 & 0.9612 & 1.5803 & 0.4159 & +18:01:22.59 & -29:21:20.06 & $--$ & $--$\\
175628 & 9 & 5 & 4.6249 & 1.6673 & 0.2966 & +17:58:27.24 & -29:52:02.39 & $--$ & OGLE-BLG-ECL-220539*\\
305199 & 9 & 1 & 4.1276 & 0.9349 & 0.6976 & +17:59:36.48 & -28:40:50.44 & $--$ & OGLE-BLG-ECL-233520*\\
356601 & 9 & 6 & 4.1813 & 1.0998 & 0.6422 & +18:00:23.29 & -29:56:56.28 & $--$ & OGLE-BLG-ECL-242433*\\
268839 & 9 & 7 & 4.1528 & 1.6834 & 0.3196 & +18:01:08.13 & -29:38:53.95 & $--$ & OGLE-BLG-ECL-250942*\\
207541 & 9 & 9 & 7.8195 & 1.3760 & 0.1397 & +18:01:30.01 & -28:57:00.47 & $--$ & OGLE-BLG-ECL-254823*\\
314103 & 10 & 5 & 0.2182 & 1.0595 & 0.4664 & +17:59:51.53 & -28:05:07.24 & $--$ & OGLE-BLG-ECL-000106*\\
50256 & 9 & 4 & 0.1849 & 1.1874 & 0.8021 & +17:57:20.60 & -29:33:50.31 & $--$ & OGLE-BLG-ECL-207454*\\
234255 & 9 & 6 & 0.1569 & 1.3045 & 0.9545 & +18:01:17.51 & -29:57:21.00 & $--$ & $--$\\
340135 & 9 & 10 & 0.0950 & 1.4823 & 0.8985 & +18:00:23.24 & -28:23:41.14 & $--$ & OGLE-BLG-ECL-000110**\\
129173&10&1&0.5603&1.2395&0.0567&+17:59:05.20&-26:53:45.98&0.279105& OGLE-BLG-ECL-227801*\\
360325&10&7&0.2995&1.6164&0.1658&+18:00:40.33&-27:44:28.88&0.150731& OGLE-BLG-ECL-245557*\\
115233&10&9&0.3315&1.5493&0.0776&+18:02:00.84&-27:06:42.65&0.163787& OGLE-BLG-ECL-260381*\\
		\hline
	\end{tabular}
    \begin{tablenotes}
      \small
      \item * within 1 arcsec\\  ** within 2 arcsec
    \end{tablenotes}
    \label{tab:example_ebs}
\end{table*}

\begin{figure*}
   \includegraphics[width=.32\textwidth]{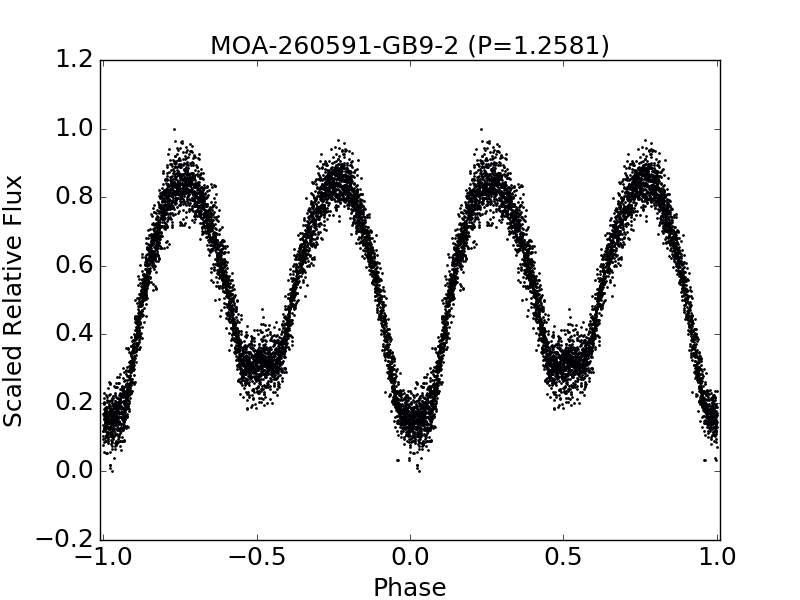}
   \includegraphics[width=.32\textwidth]{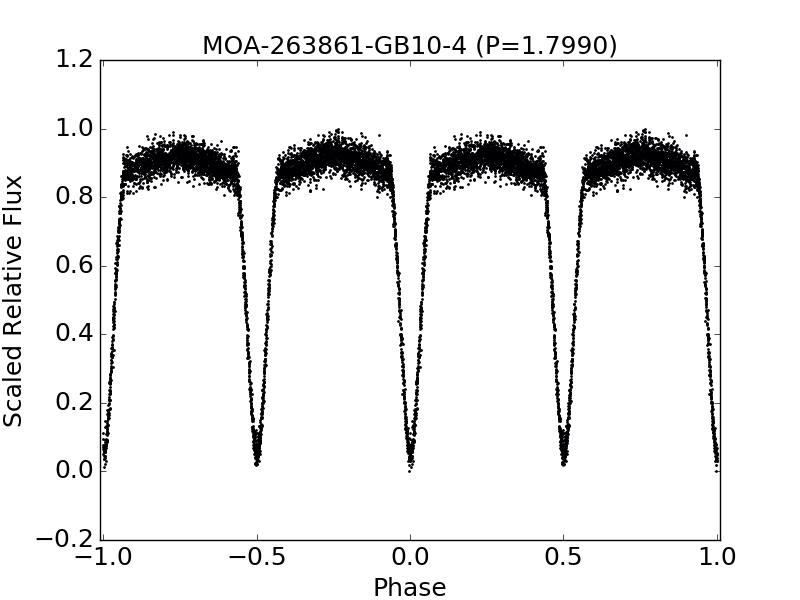}
   \includegraphics[width=.32\textwidth]{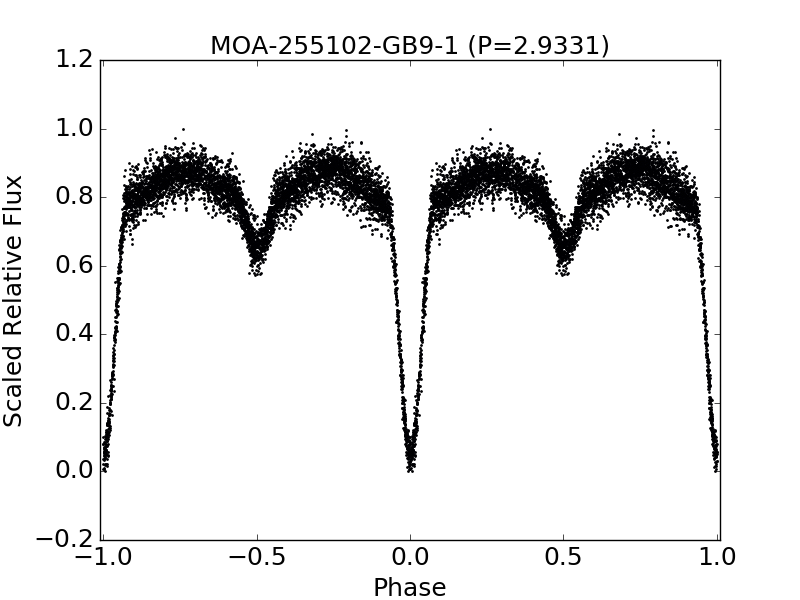}
   \includegraphics[width=.32\textwidth]{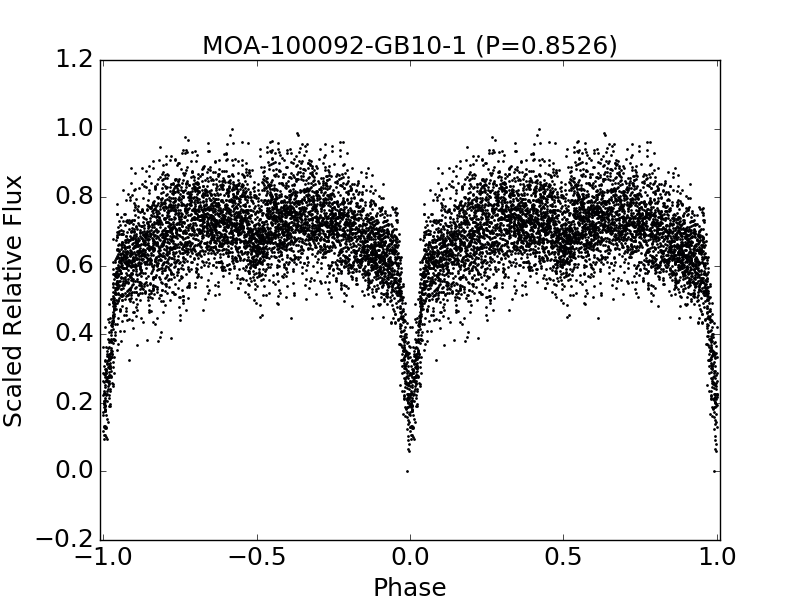}
\includegraphics[width=.32\textwidth]{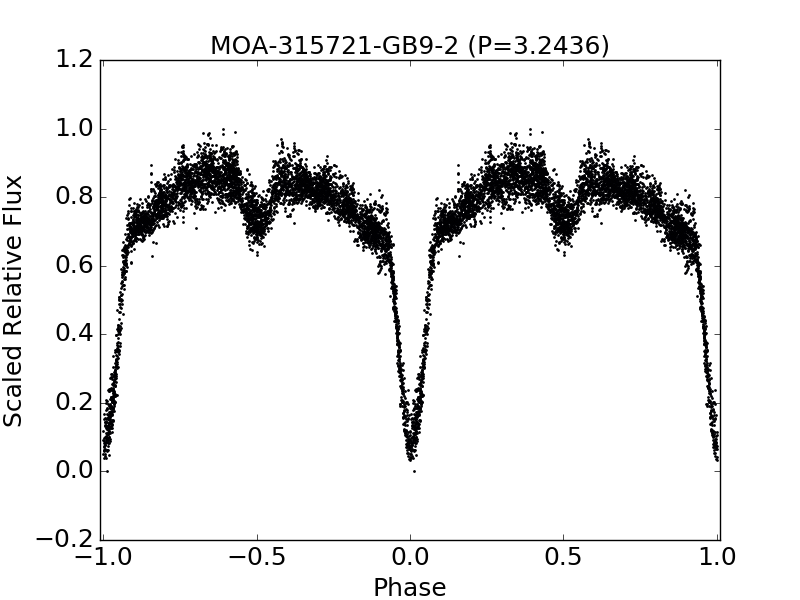}
\includegraphics[width=.32\textwidth]{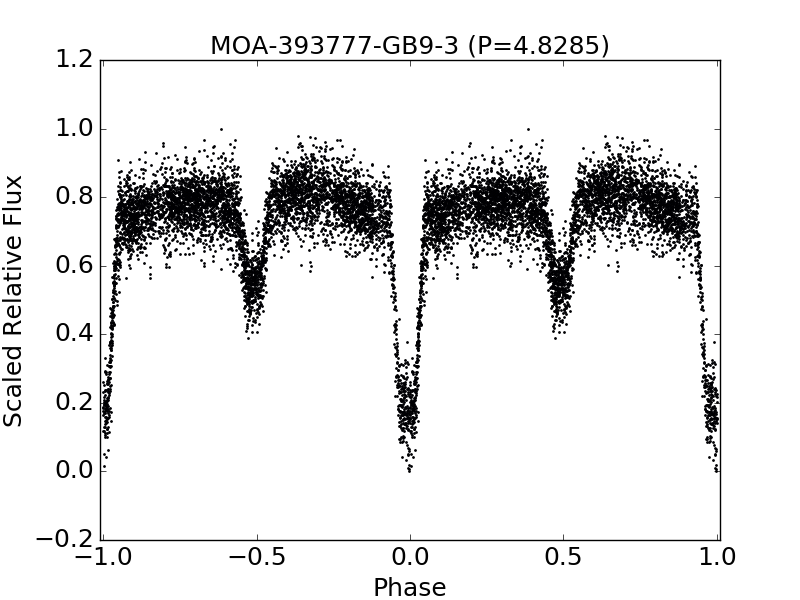}
   \includegraphics[width=.32\textwidth]{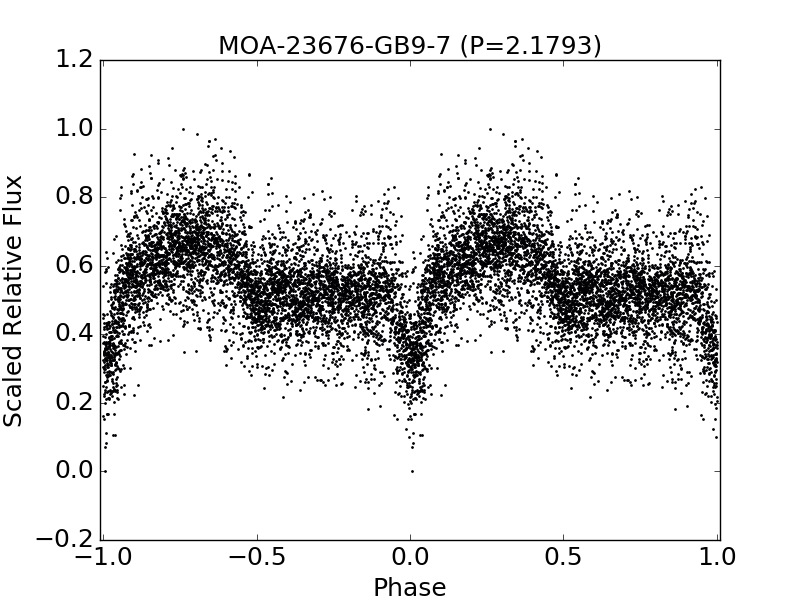}    \includegraphics[width=.32\textwidth]{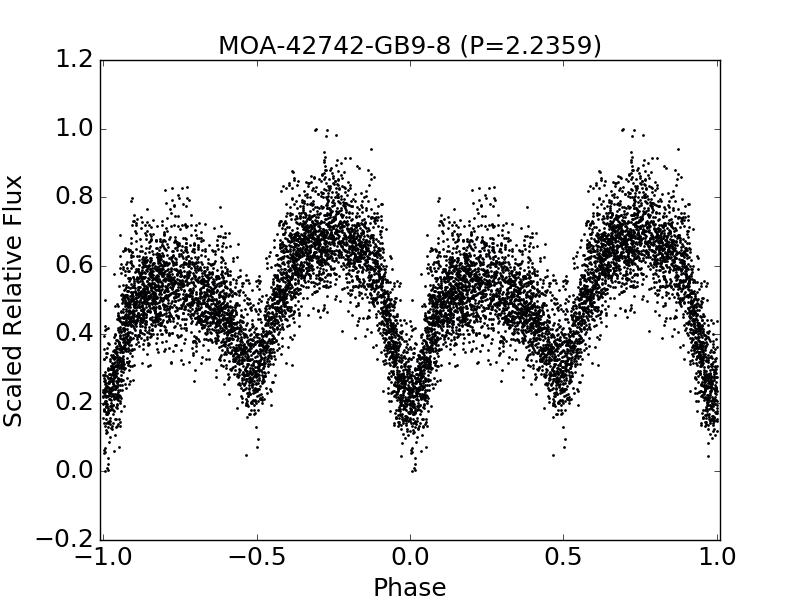}
   \includegraphics[width=.32\textwidth]{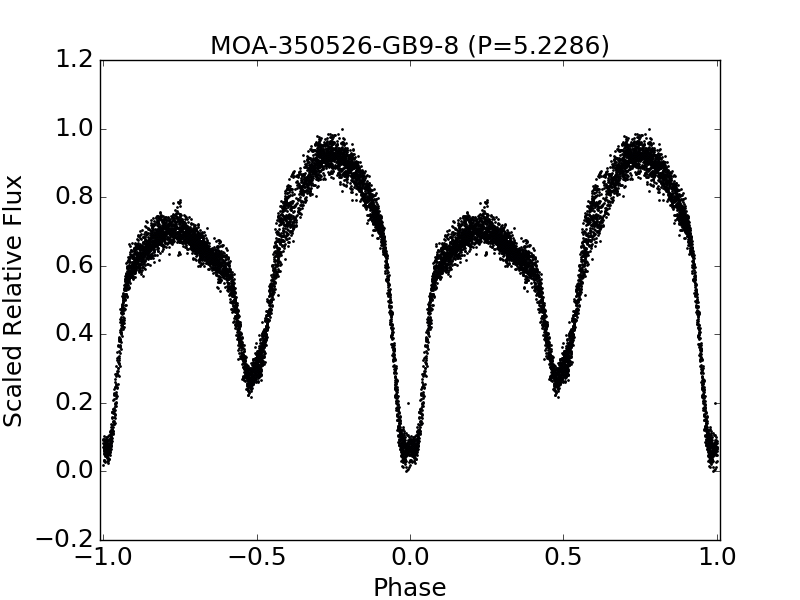}
   \includegraphics[width=.32\textwidth]{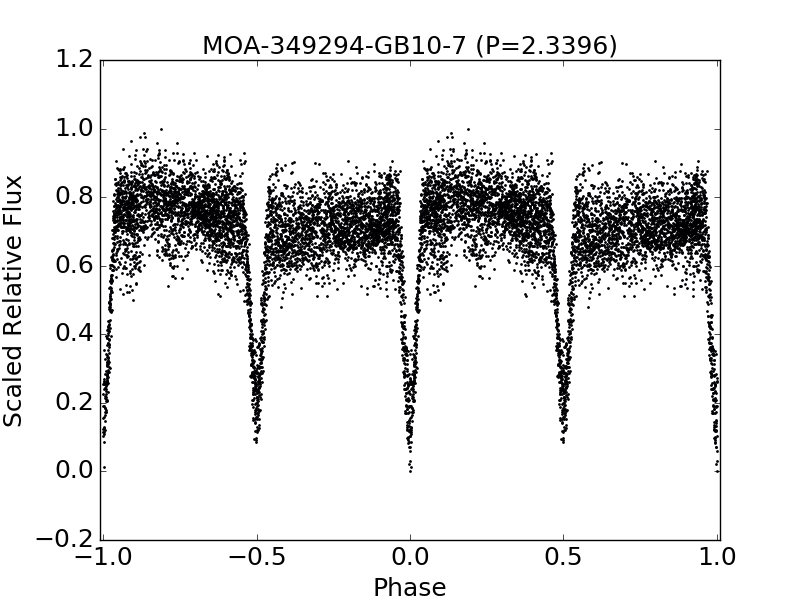}
    \includegraphics[width=.32\textwidth]{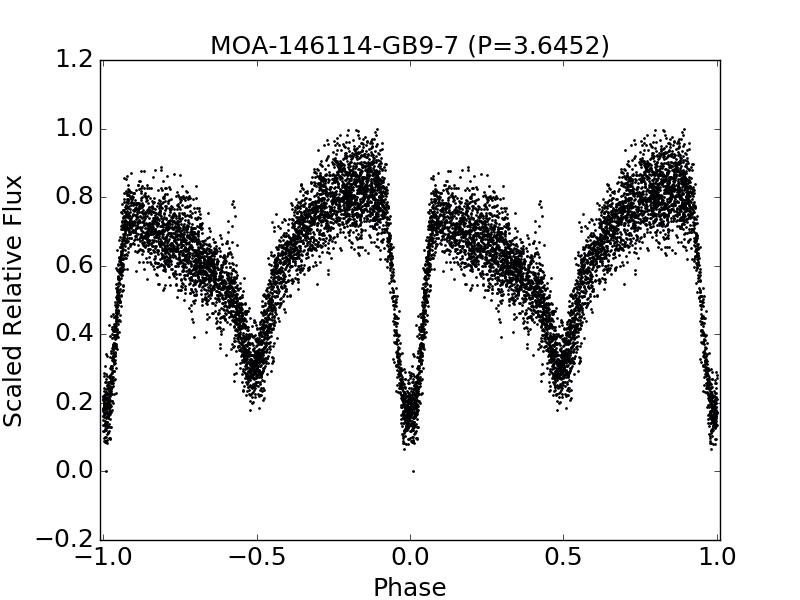}
    \includegraphics[width=.32\textwidth]{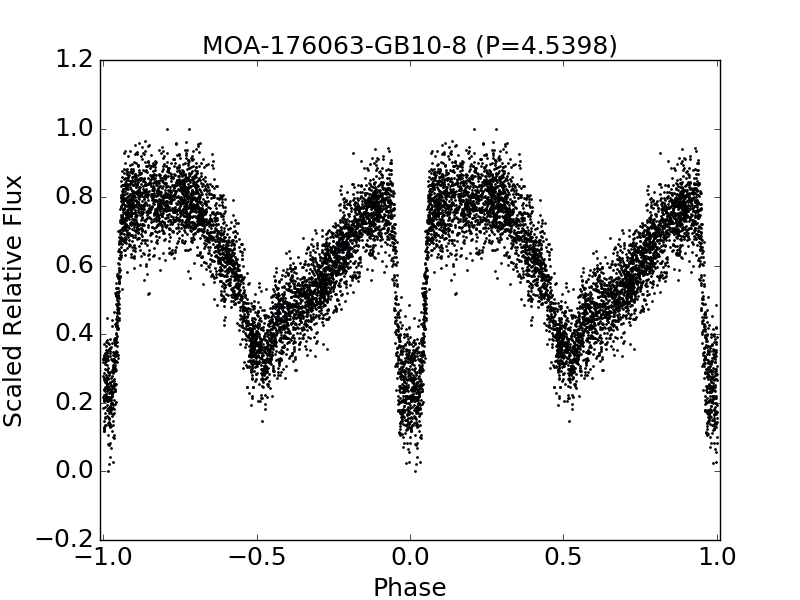}
    \includegraphics[width=.32\textwidth]{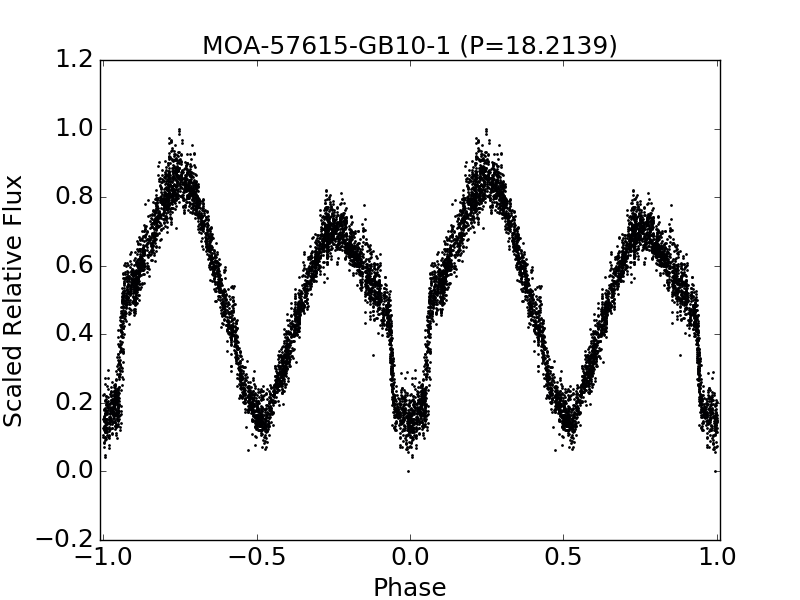}
   \includegraphics[width=.32\textwidth]{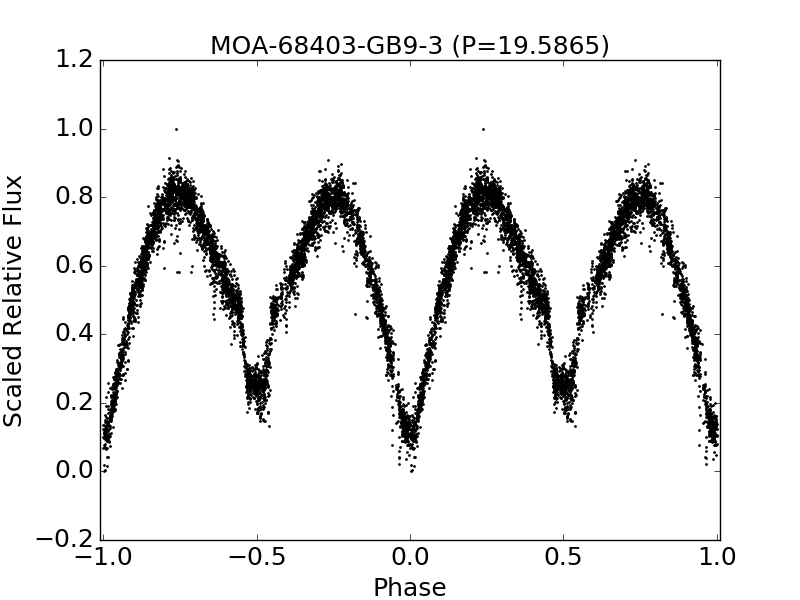}
   \includegraphics[width=.32\textwidth]{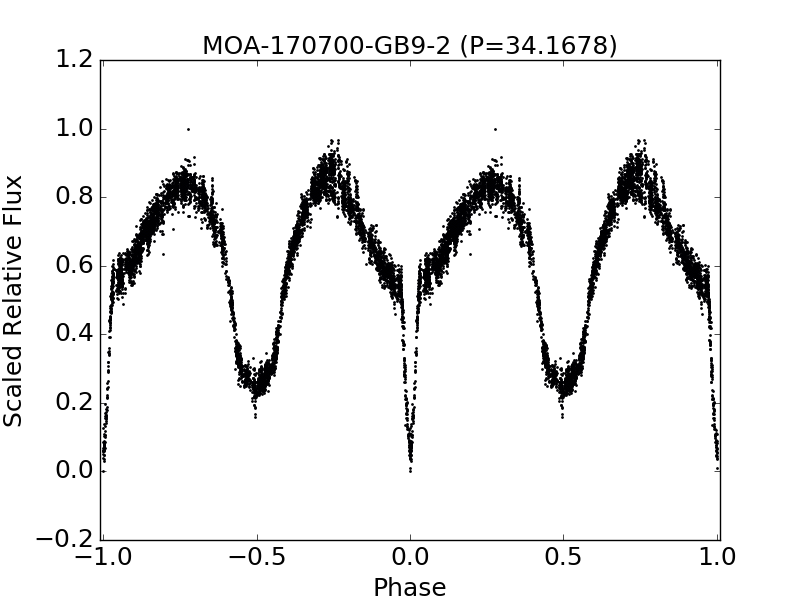}
    \caption{Examples of the folded light curves of eclipsing binaries (EBs) in the GB9 and GB10 fields, including those with dominant ellipsoidal modulations (first row), or reflection effects (second row), those showing strong O'Connell effects (third row), and the three EBs with unusual phase modulations potentially involving multiple effects in additional to reflection effect (fourth row). The bottom row shows three examples of binaries with giant or sub-giant star components. The eclipsing periods are in days. The minimum of the primary eclipse of each folded light curve was adjusted to be located at the zero phase.}
    \label{fig:example_lc}
\end{figure*}

\begin{figure*}
	\includegraphics[width=.32\textwidth]{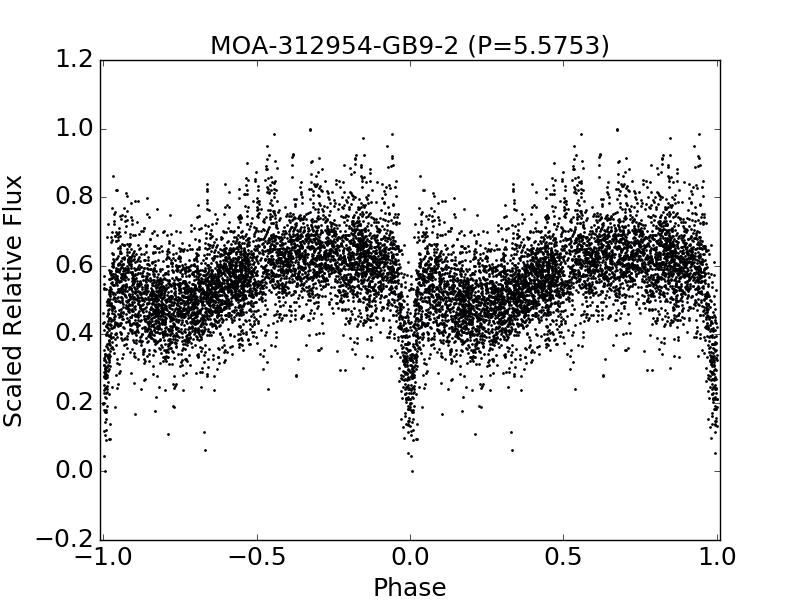}
    \includegraphics[width=.32\textwidth]{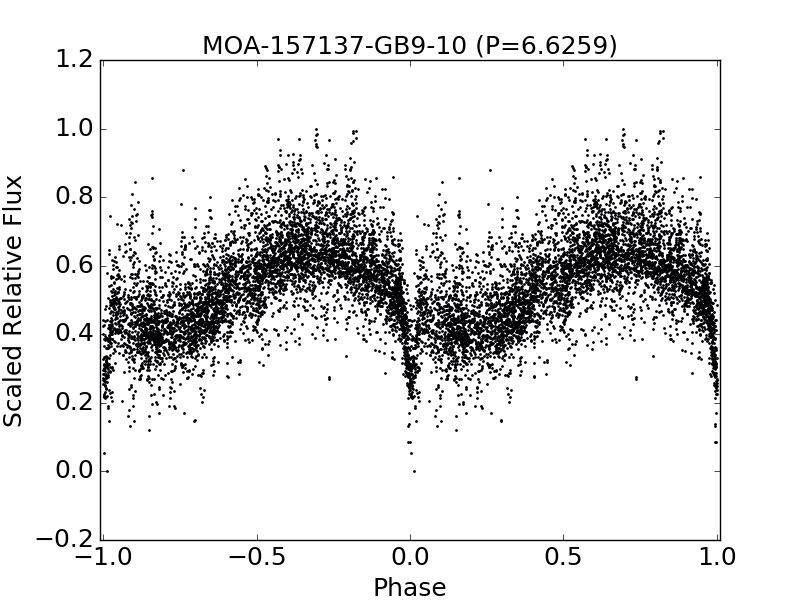}
    \includegraphics[width=.32\textwidth]{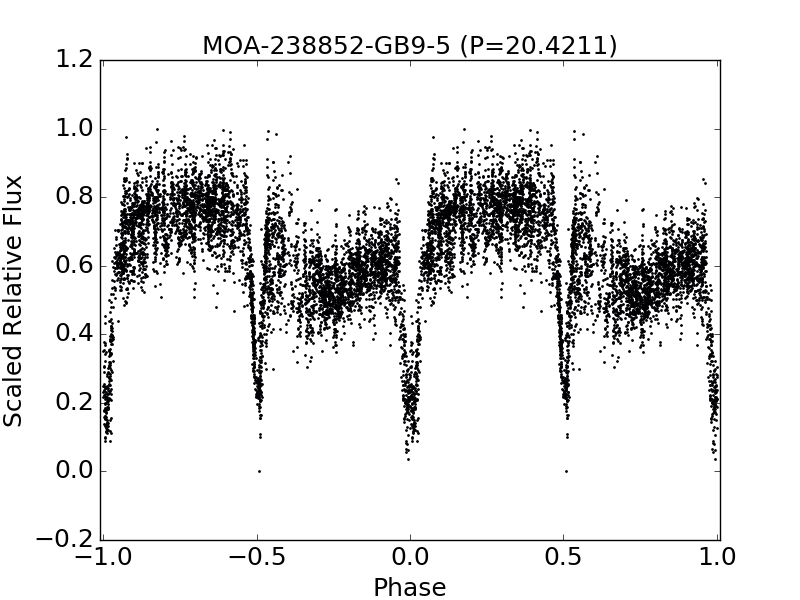}
    \includegraphics[width=.32\textwidth]{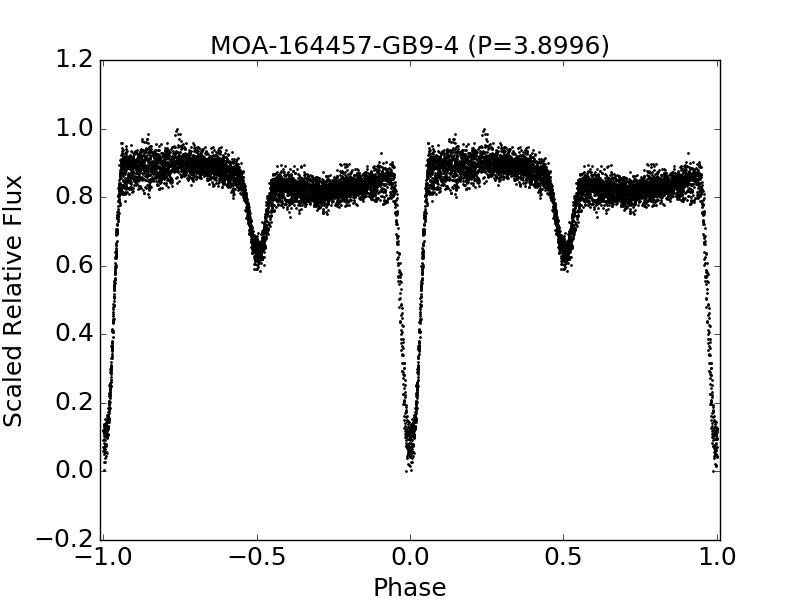}
    \includegraphics[width=.32\textwidth]{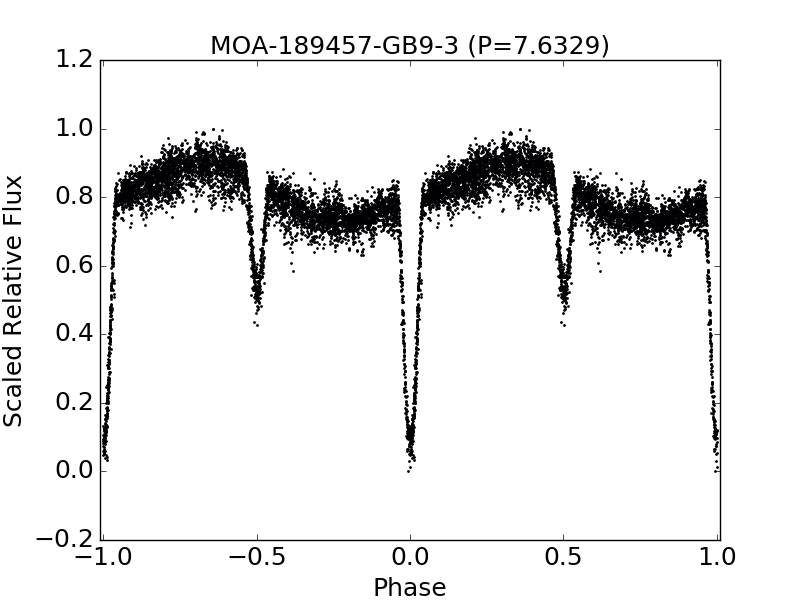}
    \includegraphics[width=.32\textwidth]{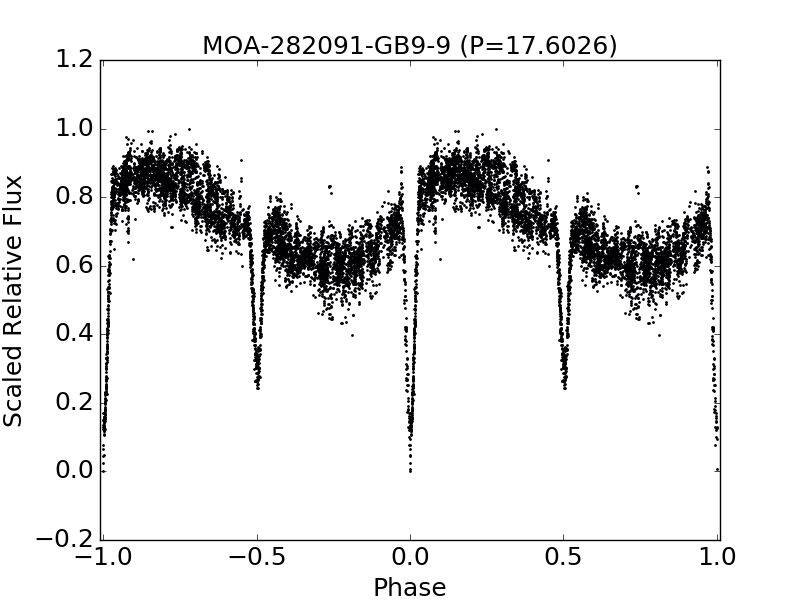}
    \caption{The folded light curves of eclipsing binaries,  in the GB9 and GB10 fields, with phase modulations potentially associated with ellipsoidal variations modified by strong reflection effects. The eclipsing periods are in days. The minimum of the primary eclipse of each folded light curve was adjusted to be located at the zero phase.}
    \label{fig:ell_mod_reflect_lc}
\end{figure*}

\begin{figure*}
    \includegraphics[width=.32\textwidth]{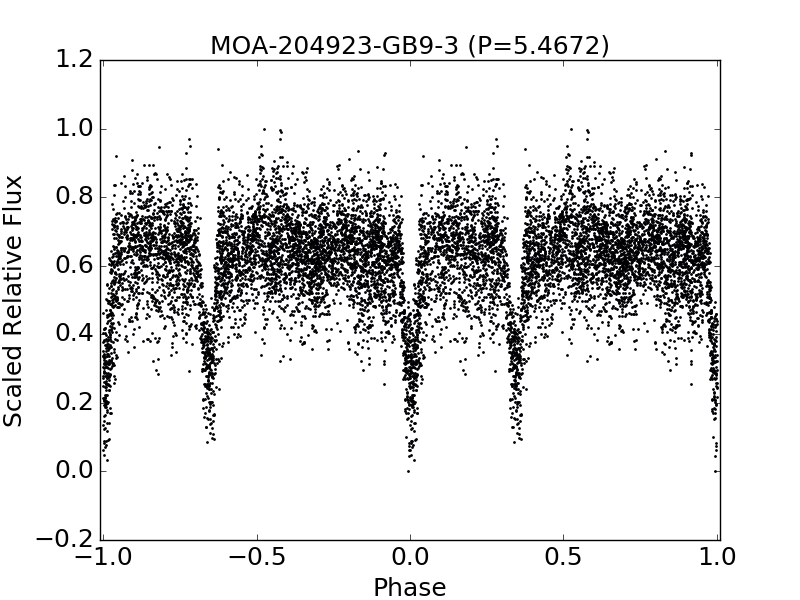}
    \includegraphics[width=.32\textwidth]{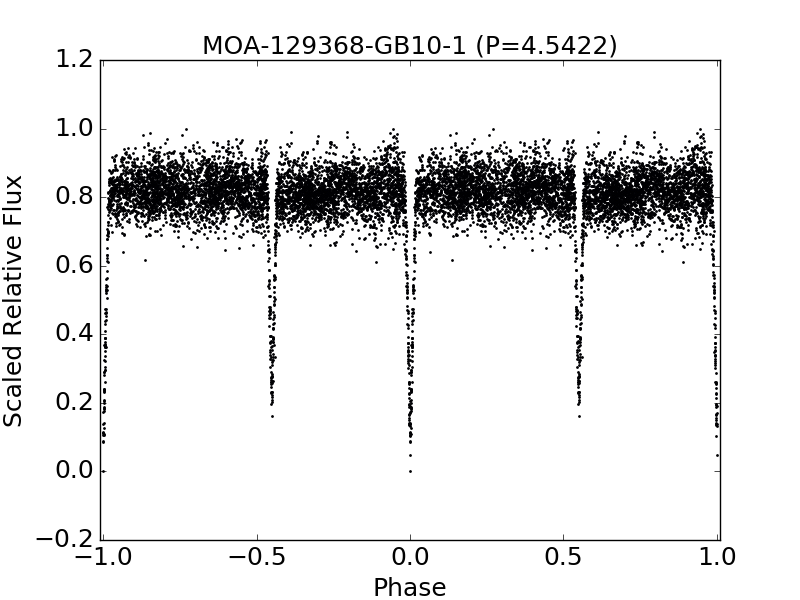}
\includegraphics[width=.32\textwidth]{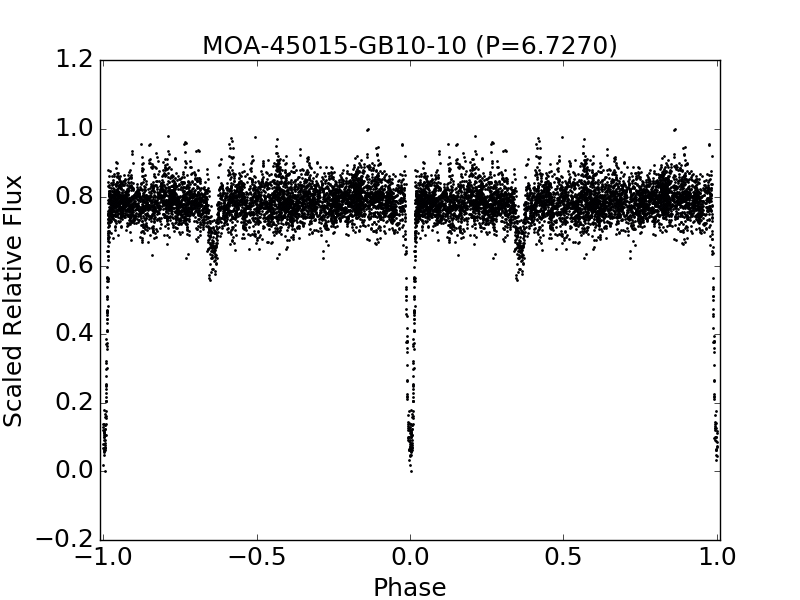}
\includegraphics[width=.32\textwidth]{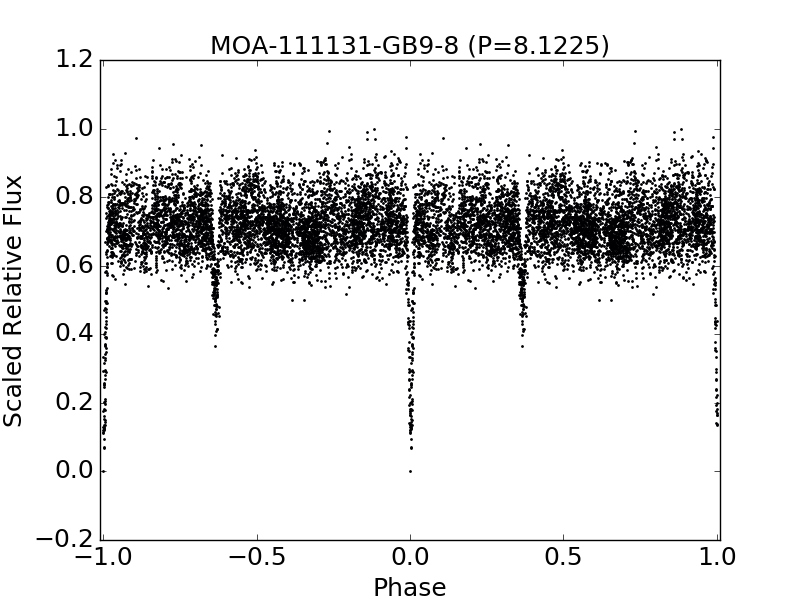}
\includegraphics[width=.32\textwidth]{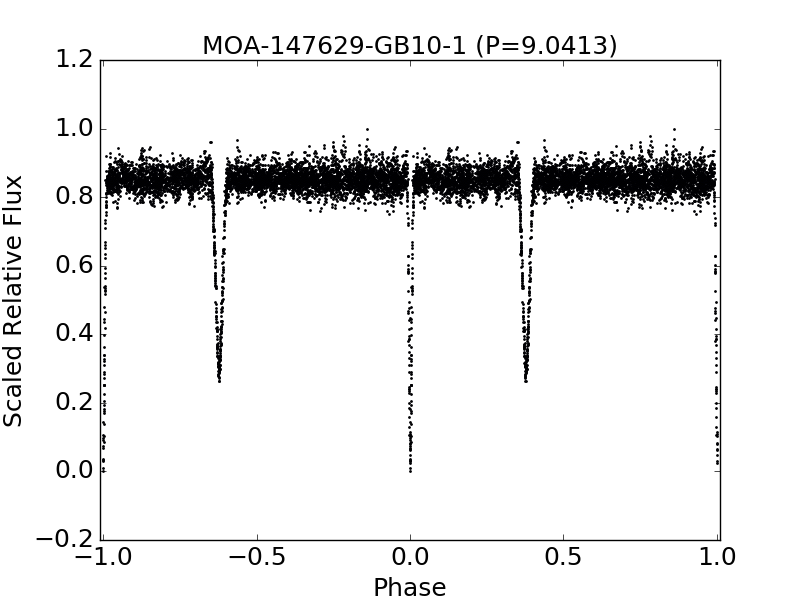}
\includegraphics[width=.32\textwidth]{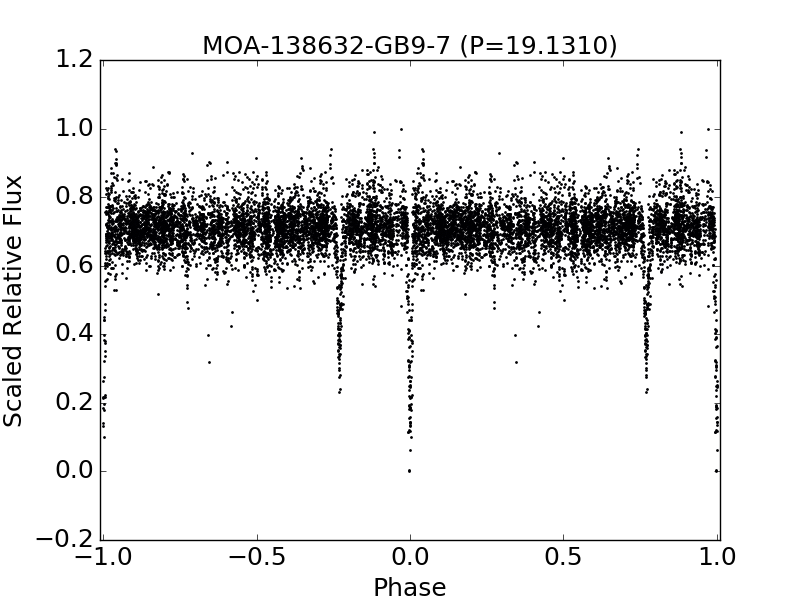}
    \caption{The folded light curves of six eccentric binaries in the GB9 and GB10 fields. The eclipsing periods are in days. The minimum of the primary eclipse of each folded light curve was adjusted to be located at the zero phase.}
    \label{fig:ecc_lc}
\end{figure*}

\section{Results}
\label{sec:results}
\subsection{Eclipsing Binary Catalogue}
Using the method described in Section~\ref{sec:iden_eb}, we identified 8,733 EB candidates with periods ranging from 0.09 day to 66 days in the GB9 and GB10 fields. Figure~\ref{fig:period_hist} shows the histogram of the EB candidates' periods with logarithmic bins, in which a cut-off in period exists at $\sim$30$\,$d (excluding the four EB candidates of periods $>40\,$d), indicating that our search was biased towards EBs with periods $<30\,\text{d}$. On the other hand, the majority of EB candidates discovered are of periods $<1\,\text{d}$ and, unsurprisingly, belong to contact or semi-detached binaries. It is manifest that detecting EBs, particularly detached binaries, of periods \(>30\,\text{d}\) is difficult using only two observational seasons worth of data. The identification rate of such long period binaries should be improved using the data with a longer time base. Since the GB9 and GB10 fields both overlap with the OGLE fields, we thus cross-checked our candidates with the OGLE collections of EBs towards the GB from the OGLE-II, OGLE-III and OGLE-IV surveys \citep{2016AcA....66..405S}. There are 486 EB candidates in our catalogue which have no OGLE EB counterpart within $4\arcsec$.

\begin{figure*}
    \includegraphics[width=.32\textwidth]{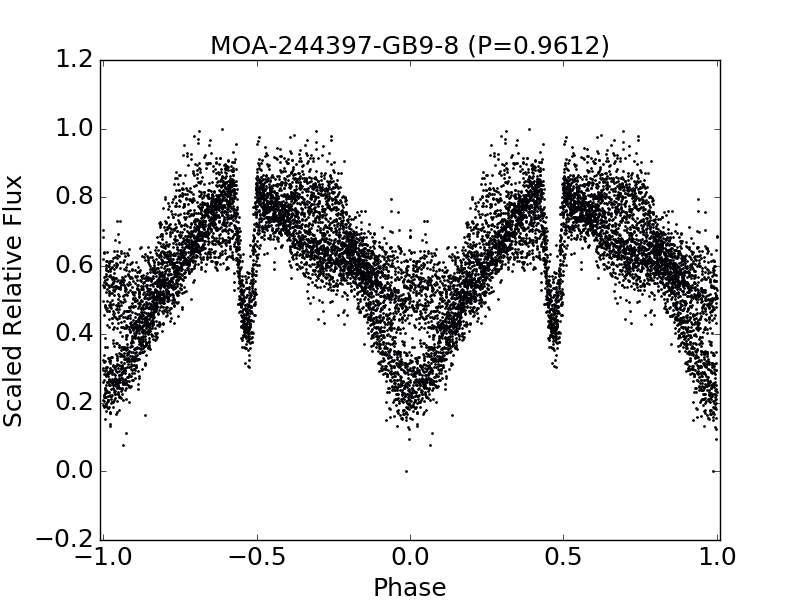}
	\includegraphics[width=.32\textwidth]{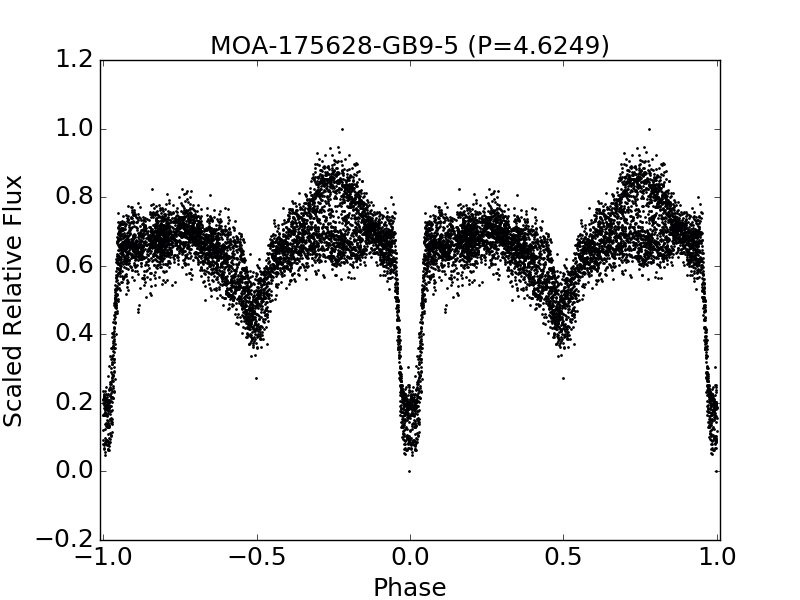}
    \includegraphics[width=.32\textwidth]{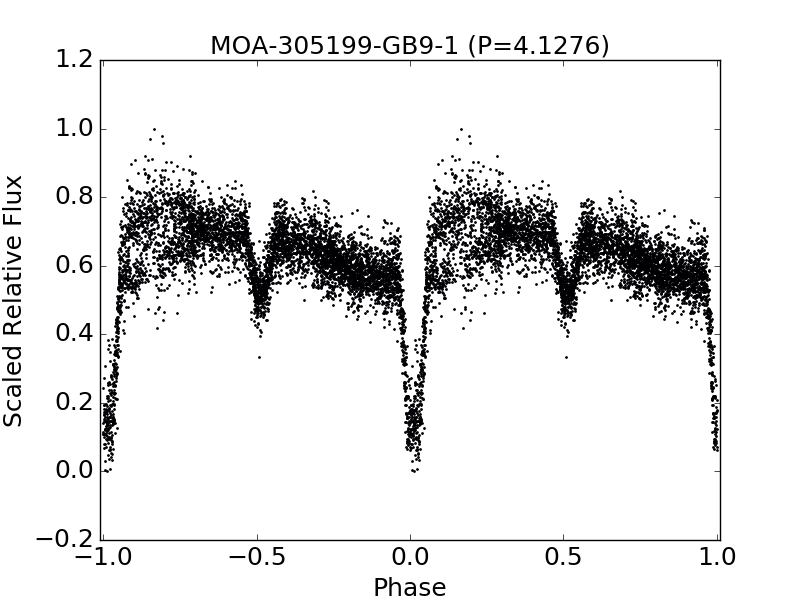}
    \includegraphics[width=.32\textwidth]{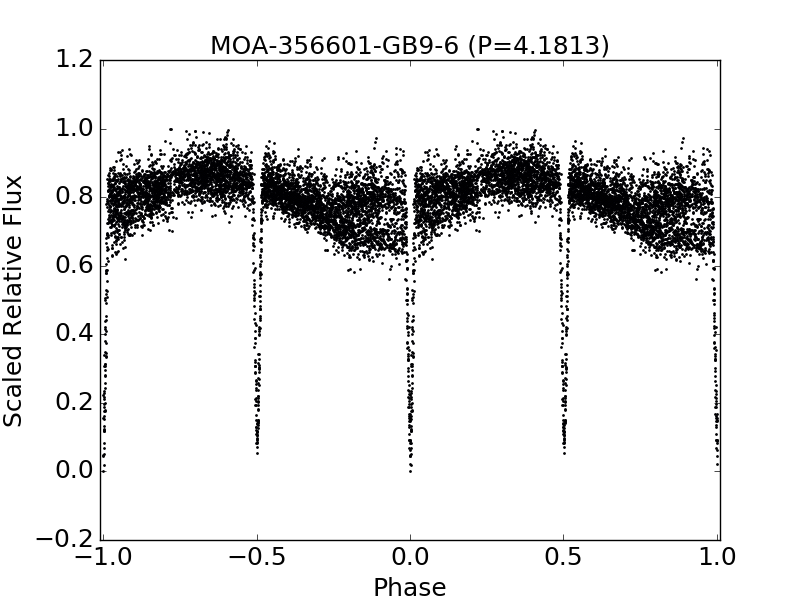}
    \includegraphics[width=.32\textwidth]{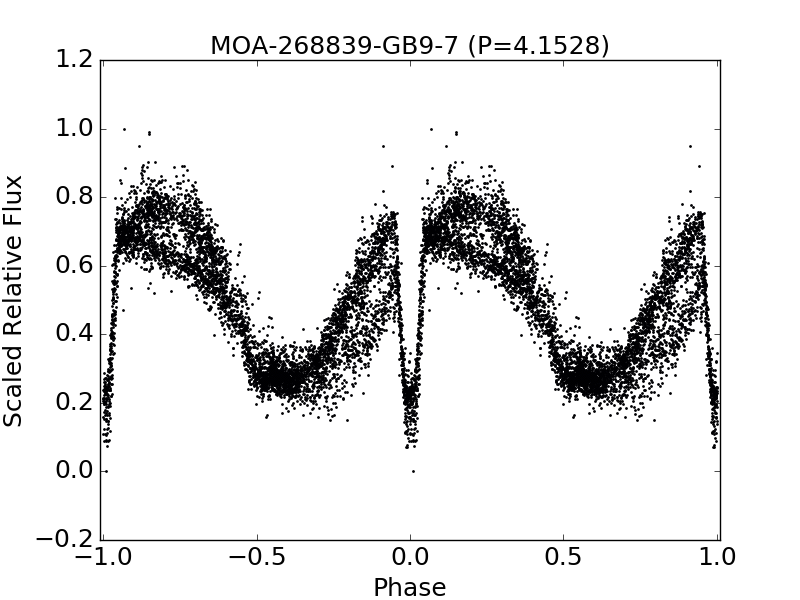}
    \includegraphics[width=.32\textwidth]{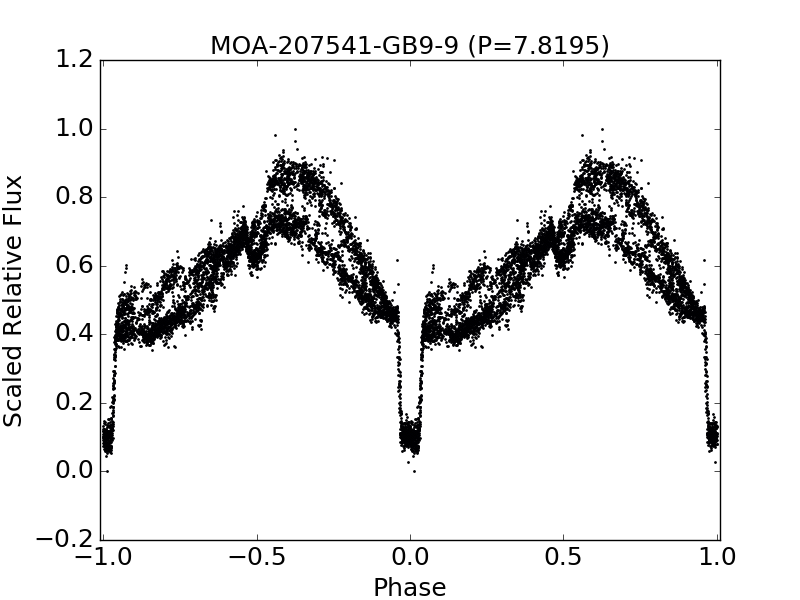}
    \caption{The folded light curves of six eclipsing RS Canum Venaticorum (RS CVn) type star candidates in the GB9 and GB10 fields. The eclipsing periods are in days. The minimum of the primary eclipse of each folded light curve was adjusted to be located at the zero phase.}
    \label{fig:ecl_RSCn_lc}
\end{figure*}


\begin{figure*}
    \includegraphics[width=\columnwidth]{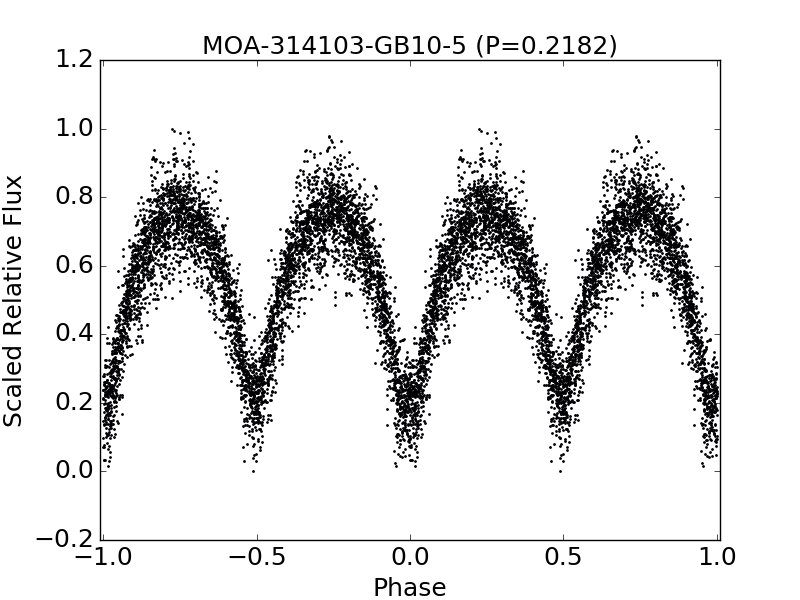}
\includegraphics[width=\columnwidth]{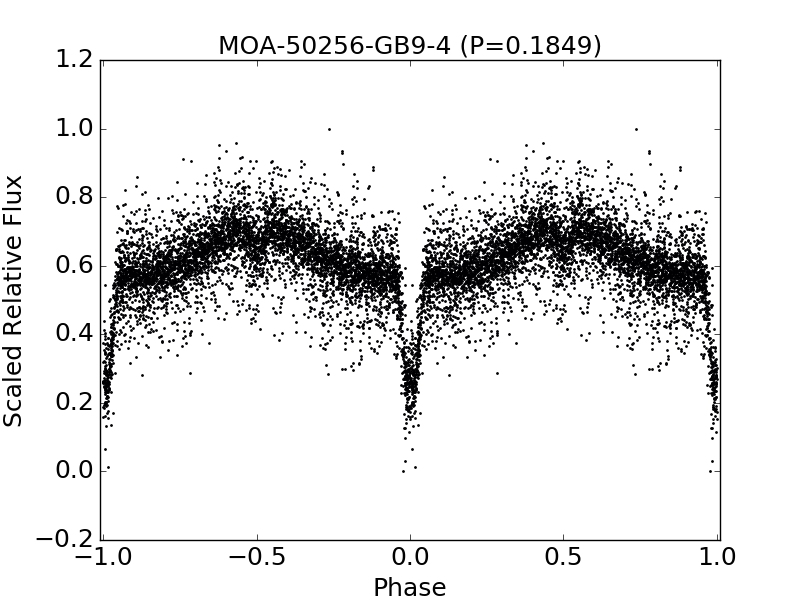}
\includegraphics[width=\columnwidth]{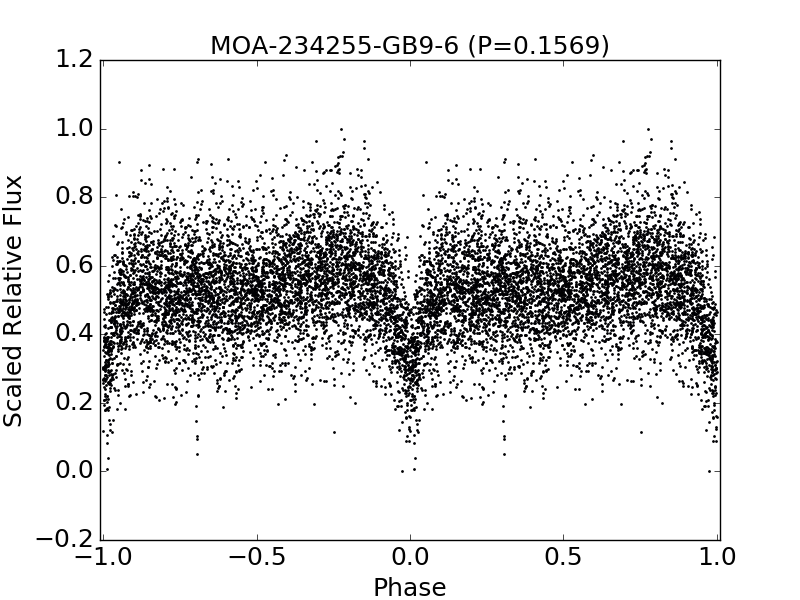}
\includegraphics[width=\columnwidth]{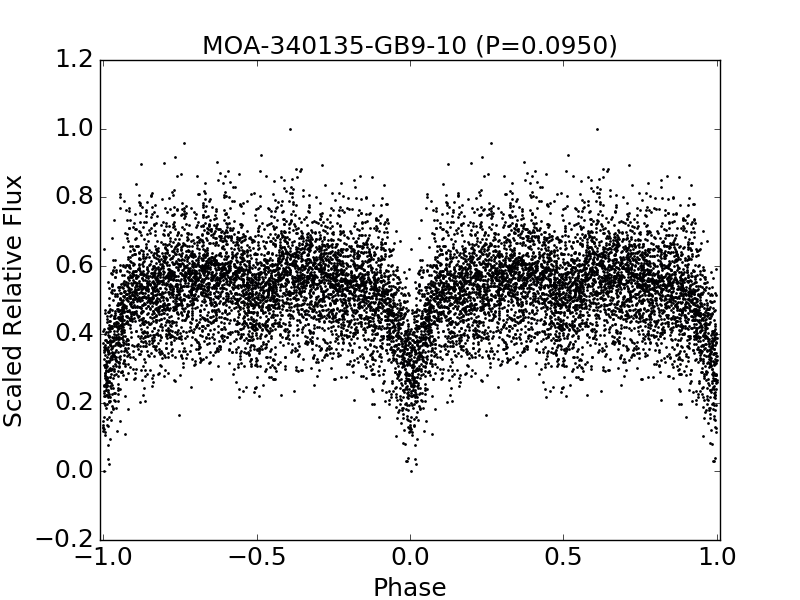}
    \caption{The folded light curves of four eclipsing binaries in the GB9 and GB10 fields with periods below the 0.22 day contact binary minimum. MOA-314103-GB10-5 is clearly a contact binary and MOA-50256-GB9-4 is a post-common-envelope binary candidate consisting of a main-sequence star and a subdwarf, judging from their light curve morphology, while the identities of MOA-234255-GB9-6 and MOA-340135-GB9-10 are unknown. Note that the eclipsing periods are in days. The minimum of the primary eclipse of each folded light curve was adjusted to be located at the zero phase.}
    \label{fig:ultrashort_lc}
\end{figure*}

\subsubsection{Eclipsing Binaries with Various Phase Modulations}
\label{sec:phase_mod_EBs}
In spite of lacking long period binaries, our catalogue already contains a large variety of EBs in terms of the light curve morphology. Many of them have phase modulations obviously dominated by ellipsoidal variations due to stellar surface distortion by tidal interaction \citep{1976ApJ...203..182W} (e.g. MOA-263861-GB10-4 and MOA-255102-GB9-1 in Figure~\ref{fig:example_lc}) or dominated by reflection effects \citep{1990ApJ...356..613W} (e.g. MOA-100092-GB10-1 and MOA-315721-GB9-2 in Figure~\ref{fig:example_lc}). The O'Connell effect \citep{1984ApJS...55..571D} is observed in a large number of ellipsoidal EB candidates, amongst which a few extreme cases were discovered (e.g. MOA-23676-GB9-7, MOA-42742-GB9-8, and MOA-350526-GB9-8 in Figure~\ref{fig:example_lc}). Also, we noticed certain numbers of EB candidates having unusual phase modulations to which multiple effects might take significant contributions in additional to the reflection effect (see Figure~\ref{fig:example_lc}). In addition, our discovery includes a group of EB candidates showing similar sinusoidal phase modulations, where the conjunction phases, in which eclipses would occur, happen to be midway between the maxima and minima of the sinusoidal curves, potentially resulting from ellipsoidal modulations modified by strong reflection effects (e.g. \citealt{2016A&A...592A..32L}) (see Figure~\ref{fig:ell_mod_reflect_lc}).

\subsubsection{Eccentric and Other Interesting Eclipsing Binaries}
\label{sec:ecc_other_EBs}
We also discovered over hundreds of detached binaries. Among them, at least six eccentric binaries were identified by inspecting their folded light curves (see Figure~\ref{fig:ecc_lc}), in which the phase differences between their own primary and secondary eclipses are noticeably different from 0.5 phase, indicating non-zero eccentricity of their orbits.

Other interesting EB candidates include those with giant or sub-giant star components (e.g. MOA-170700-GB9-2 in Figure~\ref{fig:example_lc}), PCEBs (e.g. MOA-50256-GB9-4 in Figure~\ref{fig:ultrashort_lc}), and eclipsing RS Canum Venaticorum (RS CVn) type star candidates, where their identities were deduced by their periods and the characteristics of their folded light curve shapes recognized by eye. For example, the ultra-short period plus the strong reflection modulation of MOA-50256-GB9-4, which is clearly a detached binary from the manifest ingress and egress, and the very short phase duration, of its primary eclipse, satisfy the characteristics of the typical PCEB light curve shape (e.g. \citealt{2015ApJ...808..179D}); and a group of EB candidates shown in Figure~\ref{fig:ecl_RSCn_lc} are considered as eclipsing RS CVn type stars because of the quasi-periodic brightness variations in the out-of-eclipse phase regions of their folded light curves, which are the characteristic features of RS CVn type star light curves owing to their active star spots (e.g. \citealt{2016ApJ...832..207R}).
 
\begin{table*}
	\centering
	\caption{Doubly eclipsing binaries in the GB10 field.}
	\begin{tabular}{cccccccccccc} 
		\hline
        No. & GB & chip & $P_{a}$ & $(H_{c})_{a}$ & $\Theta_{a}$ &$P_{b}$ & $(H_{c})_{b}$ & $\Theta_{b}$ & RA & Dec & OGLE Object\\
         & & &(d)& & &(d)& & &(h:m:s)&(d:m:s)& ($<4$ arcsec)\\
        \hline
        280006 & 10 & 9 & 2.6414 & 1.7097 & 0.3056 & 0.4114 & 1.3856 & 0.7136 & +18:00:45.08 & -27:18:00.50 & OGLE-BLG-ECL-246476*\\

129510& 10 & 9 & 0.7261 & 1.7515 & 0.7322 & 0.3854 & 1.4043 & 0.7620 & +18:01:53.99 & -27:09:09.53 & OGLE-BLG-ECL-259166*\\

61499 & 10 & 3 & 0.4091 & 1.3412 & 0.7962 & 1.6946 & 1.2237 & 0.8102 &+17:57:40.49 & -27:31:23.57 & OGLE-BLG-ECL-211511*\\
		\hline
	\end{tabular}
    \begin{tablenotes}
      \small
      \item * within 1 arcsec
    \end{tablenotes}
    \label{tab:doubly_ebs}
\end{table*}

\begin{figure*}
	\includegraphics[width=0.32\textwidth]{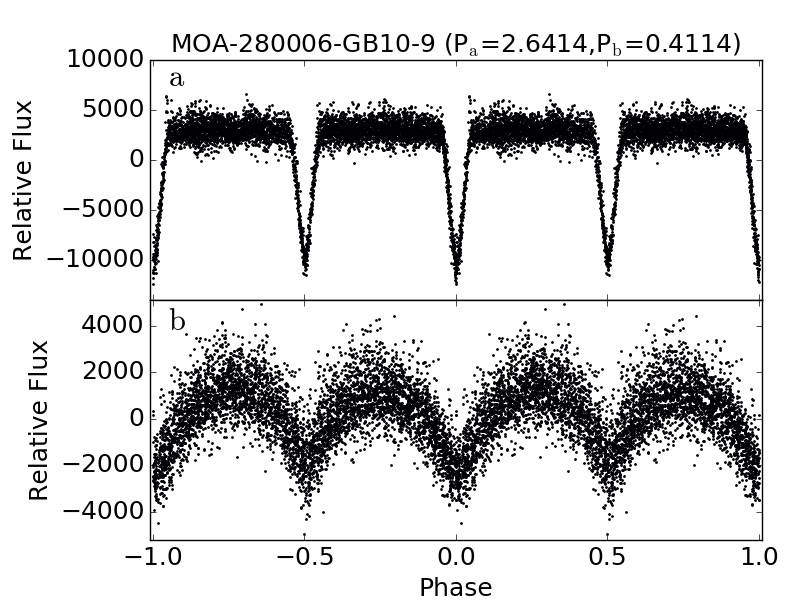}
    \includegraphics[width=0.32\textwidth]{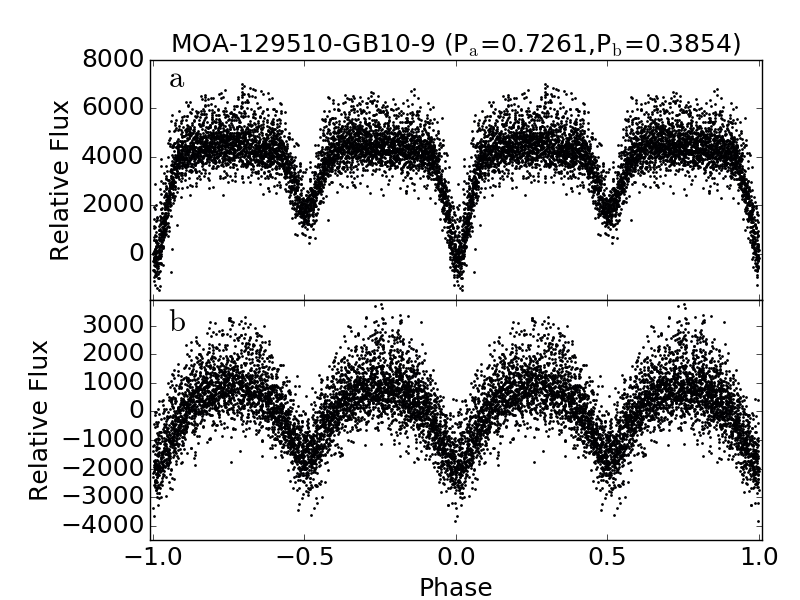}
    \includegraphics[width=0.32\textwidth]{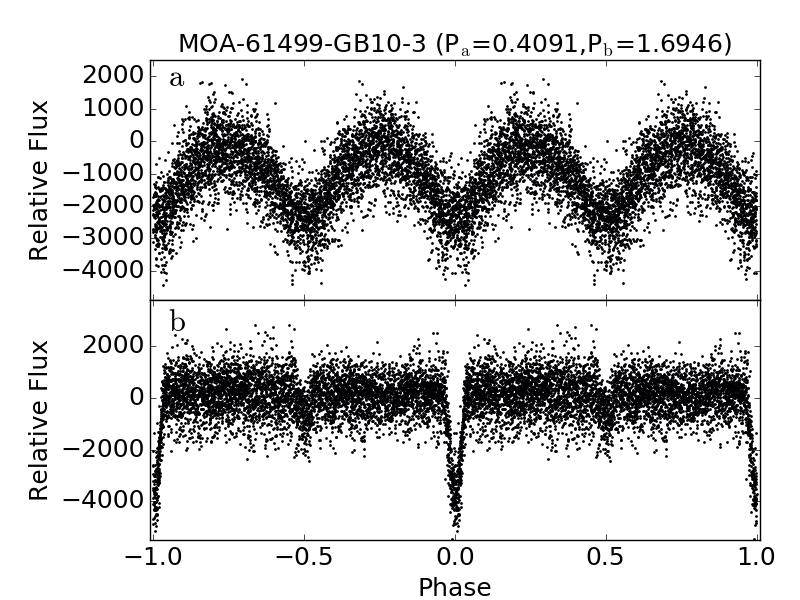}
    \caption{The folded light curves of doubly eclipsing binaries in the GB10 field. (a) The main eclipsing signals. (b) The additional eclipsing signals. Note that the eclipsing periods are in days. The minimum of the primary eclipse of each folded light curve was adjusted to be located at the zero phase.}
    \label{fig:doubly_ebs}
\end{figure*}

\subsubsection{Doubly Eclipsing Binaries}
We further searched for any hidden eclipsing signal on top of the strongest eclipsing signals as follows: we calculated the mean flux values of a folded light curve with 200 bins; we then subtracted the mean flux values from the folded light curve and unfold it afterwards; we then determined any additional eclipsing period for the subtracted light curve using the CE method, and inspected the light curves folded at the new periods. As a result, three doubly EB candidates, listed in Table~\ref{tab:doubly_ebs}, were discovered in the datasets. The folded light curves of their two eclipsing signals are shown in Figure~\ref{fig:doubly_ebs}. Given the fact that they lie in the most crowded fields, they might not be genuine but result from the blending of nearby EBs.

\subsection{Light Travel Time Effect (LTTE)}
\label{sec:ltte}
Lacking spectroscopic information, very limited study of EBs can be executed with photometric data alone. Nonetheless, it is possible to detect LTTE, which arises from a tertiary companion orbiting around an EB, by analyzing its observed-minus-calculated diagram that is constructed through the measurement of the differences between the measured and calculated eclipse times, in which the latter is obtained by extrapolation of the reference epoch of eclipse minimum assuming a constant eclipsing period, i.e.,
\begin{equation}
\Delta=T_{o}(E)-T_{c}(E)=T_{o}(E)-T_{0}-P_{s}E
\label{eq:etv}
\end{equation}
where $T_{o}(E)$ and $T_{c}(E)$ denote the observed and calculated times of the $E$-th eclipse, $T_{0}$ represents the reference epoch and $P_{s}$ denotes the eclipsing period.

\subsubsection{Light Curve Preparation For ETV Measurements}
As mentioned above all MOA light curves provided were recorded in JD. To detect any LTTE we must convert the times in JD to BJD in order to eliminate spurious ETVs from the heliocentric motion of the Earth as well as the Sun's motion around the barycentre of the Solar system and remove the systematic time errors arising from the time differences between UTC and TDB. In addition, we further cleaned the EB light curves by the following procedure:
\begin{enumerate}
\item  Fold the light curve at the eclipsing period, $P_{s}^{'}$, obtained in Section \ref{sec:pa},
\item  Bin the folded light curve into 200 bins in phase,
\item  Calculate the weighted mean and standard deviation of flux in each bin, and
\item Remove the data points 3 weighted standard deviation above or below the weighted flux mean.
\end{enumerate}
After this additional cleaning process, the eclipsing periods were recalculated using the CE method by minimizing the conditional entropy over trial periods within $P_{s}^{'}\pm 0.05$. The recalculated eclipsing periods were used as the input periods into the ETV analysis described below. Data points with errors $>2000$ flux units were rejected as being unreliable measurements.

\subsubsection{Measuring Eclipse Times}
\label{sec:measure_t_ecl}
To deal with the sparseness of data points covering an eclipse, the measurement of eclipse times was performed by the template method, in which a template was generated by fitting the phenomenological EB light curve model to the eclipsing portion of a folded light curve. To do this we used \texttt{emcee}, a Python implementation of the affine-invariant ensemble sampler for Markov chain Monte Carlo (MCMC) \citep{2013PASP..125..306F}. The phenomenological model we used to create the template of an eclipse was proposed by \cite{2015A&A...584A...8M}, in which the template model is defined by:
\begin{equation}
f(t_{i},\mathbf{\theta})=\alpha_{0}+\alpha_{1}\psi(t_{i},t_{0},d,\Gamma)
\end{equation}
where \(\alpha_{0}\) is the magnitude zero-point shift, i.e. the relative flux baseline level in our study, and \(\alpha_{1}<0\) is a negative multiplication constant of eclipse profile function, i.e.,
\begin{equation}
\psi(t_{i},t_{0},d,\Gamma)=1-\bigg\{1-\exp\bigg[1-\cosh\bigg(\frac{t_{i}-t_{0}}{d}\bigg)\bigg]\bigg\}^{\Gamma}.
\end{equation}
Note that \(t_{0}\) is the time of the minimum of an eclipse, $d>0$ is the minimum width and $\Gamma>0$ is the parameter specifying the pointedness of the minimum such that $\Gamma>1$ corresponds to the flat minimum associated with a total eclipse. The portions corresponding to primary and secondary eclipses were determined by estimating the minima of the second derivative of the folded light curve, which would correspond to the ingress and egress phases of the eclipses. We then fit each eclipse having at least four data points that are distributed across the eclipse minimum with the template, using \texttt{emcee}.  In this fitting we allowed only $t_{0}$, $\alpha_{0}$ and $\alpha_{1}$ to vary. In terms of the template method, the reference epoch, $T_{0}$, in eq.(\ref{eq:etv}), was defined by $T_{0}=\phi_{0}P_{s}+\tau_{0}$, where $\phi_{0}$ is the phase of the minimum of the eclipse template with respect to the time zero, $\tau_{0}$. The time zero was taken such that $\tau_{0}<t_{\text{obs}}$, where $t_{\text{obs}}$ is an observation time, and both primary and secondary eclipses are not broken in phase when the light curve is folded with respect to $\tau_{0}$. The time of an eclipse minimum is the median of the projected posterior on $t_{0}$. The uncertainty in $t_{0}$ was taken as the 1-$\sigma$ confidence interval from the median.

\begin{table*}
	\centering
	\caption{The orbital elements from the light-travel-time effect solutions for three selected eclipsing binaries in our catalogue with the uncertainties given by 1-$\sigma$ confidence intervals from the medians. Note that $P_{1}$ is the period of the inner binary determined by the conditional entropy method plus the correction, $c_{1}$, given by the best fit of eq.(\ref{eq:ltte}) to the eclipse time variation curve, and $\Delta P_{1} = 2c_{2}$, where $c_{2}$ is the second order coefficient in eq.(\ref{eq:ltte}), is the change in inner binary orbital period per orbital cycle in units of [day/cycle], and $m_\text{AB}$ was taken as $2\,M_{\sun}$ when calculating ($m_{\text{C}}$)$_{\text{min}}$.}
    \bgroup
    \def\arraystretch{1.4}
	\begin{tabular}{ccccccccccc} 
		\hline
        No. & GB & chip &$P_{1}$ &$\Delta P_{1}$& $P_{2}$ & $e_{2}$ & $\omega_{2}$ & $\tau_{2}$ & $a_{\text{AB}}\sin i_{2}$ & ($m_{\text{C}}$)$_{\text{min}}$\\
        	& & & $\times 10^{-6}\,$(d)& $\times10^{-10}\,$(d/c)& $\times 10^{-2}\,$(d) &	$\times 10^{-2}$ &(deg)& (MBJD) &(AU) & ($M_{\sun}$)\\
        \hline
        129173 &10 & 1 & 560311.71$_{-0.93}^{+0.87}$ & 58.3$_{-15.4}^{+16.6}$ &24651.1$_{-0.7}^{+0.7}$ & 20.6$_{-3.7}^{+3.5}$ & 261.6$_{-10.6}^{+9.5}$ & 56322.80$_{-0.08} ^{+0.08}$ & 0.302$_{-0.007}^{+0.007}$ & 0.76\\
        360325 & 10 & 7 & 299511.32$_{-0.62}^{+0.67}$ & 9.2$_{-6.4}^{+5.9}$&48180.7$_{-2.6}^{+1.1}$ & 25.3$_{-4.9}^{+4.4}$ & 191.9$_{-11.5}^{+8.4}$ & 56323.61$_{-0.19}^{+0.15}$ & 0.941$_{-0.038}^{+0.032}$ & 1.97\\
        115233 & 10 & 9 & 331521.24$_{-0.63}^{+0.66}$ & 5.6$_{-6.6}^{+6.3}$ &42678.8$_{-1.7}^{+1.9}$ & 30.3$_{-5.2}^{+5.4}$ & 306.1$_{-12.3}^{+13.2}$ & 56323.57$_{-0.10}^{+0.11}$ & 0.391$_{-0.015}^{+0.014}$ & 0.65\\
		\hline
	\end{tabular}
    \egroup
    \label{tab:triple}
\end{table*}

\begin{figure*}
    \includegraphics[width=0.325\textwidth]{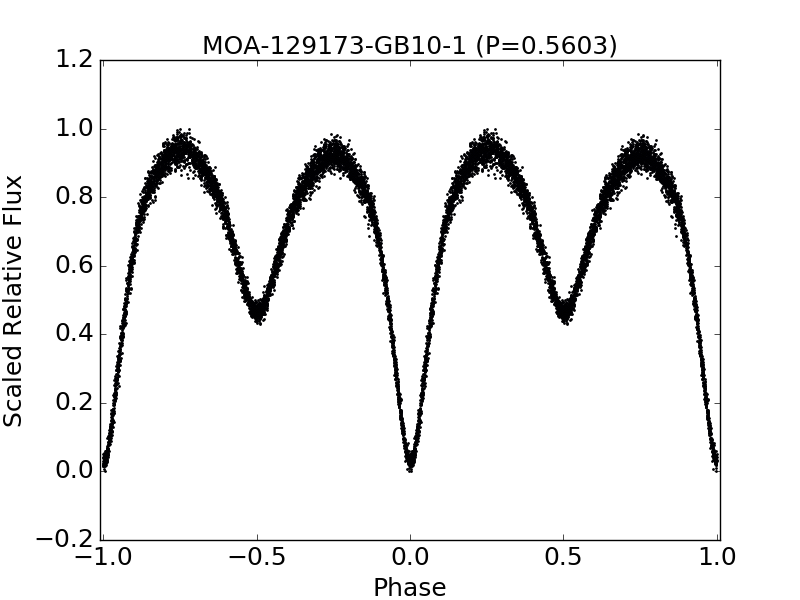}
    \includegraphics[width=0.325\textwidth]{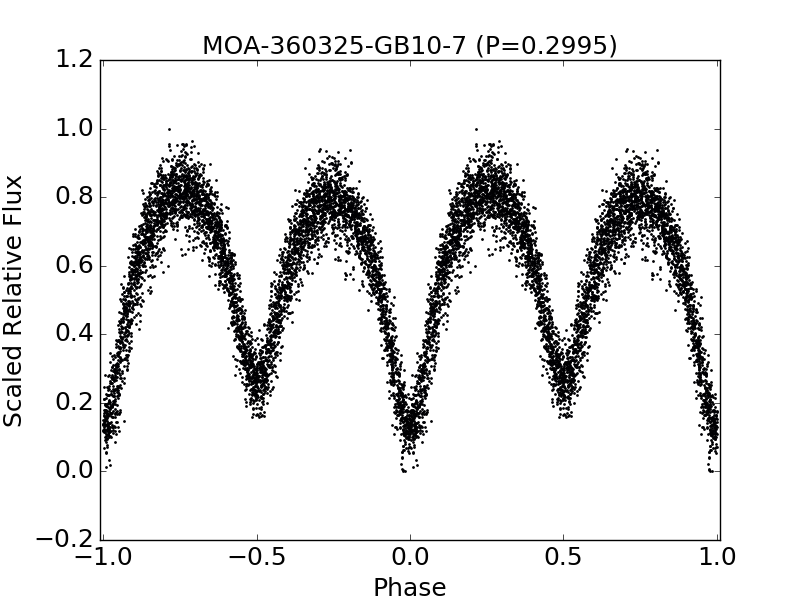}
    \includegraphics[width=0.325\textwidth]{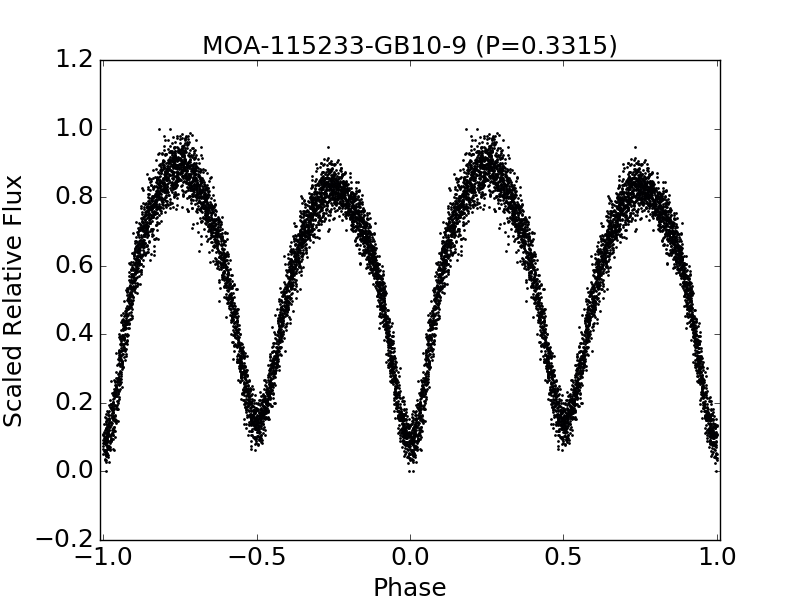}
    \includegraphics[width=0.325\textwidth]{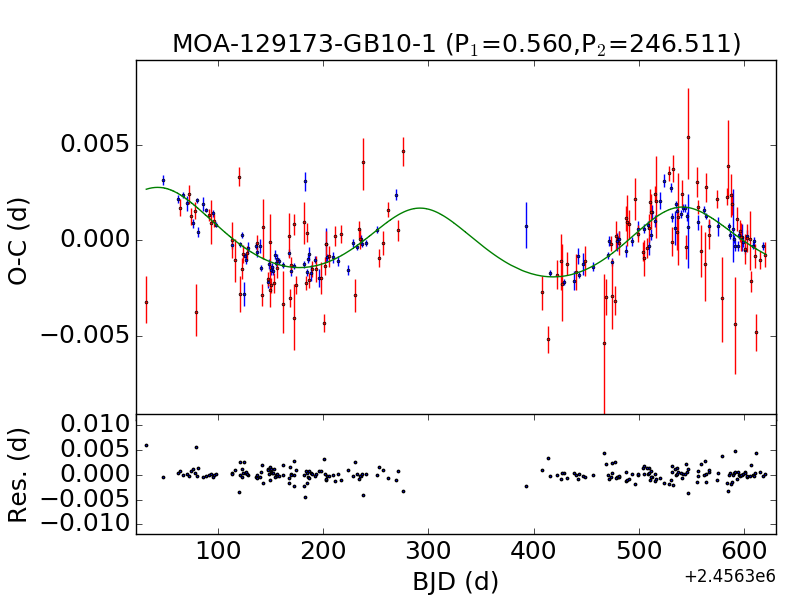}
    \includegraphics[width=0.325\textwidth]{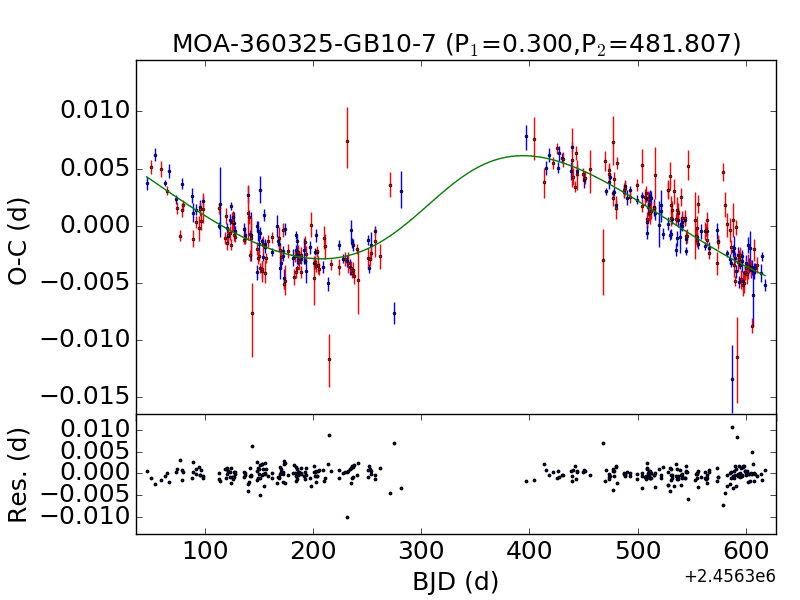}
    \includegraphics[width=0.325\textwidth]{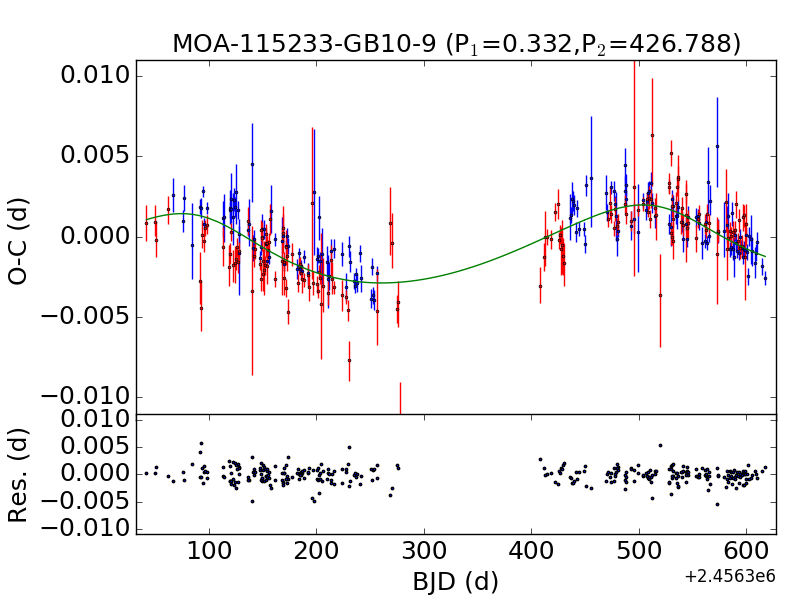}
    \caption{(Top) The folded light curves of three eclipsing binary (EB) candidates in our catalogue with detected light-travel-time effects (LTTEs); the minimum of the primary eclipse of each folded light curve was adjusted to be located at the zero phase. (Bottom) The eclipse time variation (ETV) curves of these three EB candidates; $P_{1}$ is the period of the inner binary determined by the conditional entropy method, while $P_{2}$ is the period of the tertiary companion given by the LTTE solution; the blue points are the ETV measurements of the primary eclipses and the red points are those of the secondary eclipses; the green lines represent the best fits of the ETV model defined by eq.(\ref{eq:ltte}); the bottom panels show the residuals of the fits. Note that the periods are in days.}
    \label{fig:oc_ebs}
\end{figure*}

\subsubsection{LTTE Model Fitting}
We then attempted to construct the observed-minus-calculated (O$-$C) diagrams for every EB candidate we identified, and searched for those showing promising periodic signals in their O$-$C curves by eye. As a result, we picked three EB candidates, i.e. MOA-115233-GB10-9, MOA-129173-GB10-1 and MOA-360325-GB10-7, which have low values of PDM statistic corresponding to sharp eclipsing signals and show obvious periodic patterns in their ETV curves, and fitted their O$-$C curves by the ETV model in \cite{2015MNRAS.448..946B} defined by:
\begin{equation}
	\Delta=c_{0}+c_{1}E+c_{2}E^{2}-\frac{a_{\text{AB}}\sin\\i_{2}}{c}\frac{(1-e_{2}^{2})\sin(\nu_{2}+\omega_{2})}{1+e_2\cos\nu_{2}},
    \label{eq:ltte}
\end{equation}
where the zeroth and first order coefficients, $c_{0}$ and $c_{1}$, in the polynomial of $E$ provide the corrections in $T_{0}$ and $P_{s}$, respectively, while the second order coefficient, $c_{2}$, is equal to half the rate of change in period, regardless of its origin. The parameters in the LTTE term, i.e. the last term in eq.(\ref{eq:ltte}), include the period ($P_{2}$), eccentricity ($e_{2}$), true anomaly ($\nu_{2}$), argument of periastron ($\omega_{2}$), inclination ($i_{2}$) and, implicitly, the time of periastron ($\tau_{2}$), of the tertiary object orbiting around an EB as well as the semi-major axis of its absolute orbit, $a_{\text{AB}}$, equal to $(m_{\text{C}}/m_{\text{ABC}})\,a_{2}$. Note that $m_{\text{C}}$ is the mass of the tertiary object, $m_{\text{ABC}}$ is the total mass of the triple system, and $a_{2}$ is the semi-major axis of the tertiary object's orbit around the EB and $c$ is the speed of light.

The selected EB candidates all turned out to be contact binaries after inspecting their folded light curves. As pointed out in \cite{2013ApJ...774...81T}, star spots, which are common on contact binaries, on an EB can produce spurious ETVs on primary and secondary eclipses that appear to be anticorrelated, or out of phase from each other, in the O$-$C diagram. In general, averaging the ETVs of primary and secondary eclipses is required to eliminate such spurious ETVs. Unfortunately, in our EB candidate light curves, usually only one of the primary or secondary eclipse was present in a cycle. This means that averaging is not possible for most of the MOA light curves. Nonetheless, the O$-$C curves of the primary and secondary eclipses did not appear out of phase from each other for all three selected EB candidates. Thus explaining the observed ETVs as being owing to star spots is disfavoured and therefore no averaging of the ETVs for the primary and secondary eclipses was required. We, therefore, searched for the best-fitting ETV solution for the O$-$C curve by \texttt{pymc} \citep{2015ascl.soft06005F}, another Python module of MCMC fitting algorithms, taking the ETV measurements of both primary and secondary eclipse times into the parameter search at the same time. The LTTE solutions for the three selected EB candidates are shown in Table \ref{tab:triple} and their best-fitted plots are presented in Figure~\ref{fig:oc_ebs}. The uncertainty of each parameter was taken as 1-$\sigma$ confidence interval from the median. 

Significant period changes (of order of $10^{-9}$ to $10^{-10}$ day per cycle) of the three selected EB candidates are indeed detected according to the best fits of the ETV model, i.e., eq.(\ref{eq:ltte}), to their ETV curves, indicating the possibility of mass transfer. On the other hand, since the time differences between their own primary and secondary eclipse minima (see Table~\ref{tab:example_ebs}) are all close to half of their orbital periods, the eccentricity can be safely assumed to be zero, and the periodic ETVs due to apsidal motion can thus be excluded. Meanwhile, there is no periodic signal noticeable in the residual curves, so the dynamical effect which arises from the tertiary companion being close enough to the inner binary to induce additional significant perturbation on the inner binary's orbital motion \citep{2015MNRAS.448..946B} was excluded as well. We, therefore, accept the LTTE as the only explanation for the observed periodic ETVs for all three selected EB candidates.

Because of the big gaps in the datasets due to the off-season periods in which the altitude of the GB was too low to allow good seeing observations, our selection of O$-$C curves for ETV analysis was obviously biased towards the systems of outer periods $\sim$1.5$\,$yr that a complete cycle of ETV can be recognized by eye without difficulty. Indeed a large number of good eclipse time measurement points can be obtained usually for very short period EBs given the fact that the day-night cycles and weather conditions often prevented us from obtaining good coverage for eclipses lasting longer than half a day. Also, detached binaries are commonly of periods longer than a day, and the number of cycles decreases as the period increases. All these factors make the number of good eclipse time measurement points too low for a detached binary with the time base of just two MOA observational seasons to show us any obvious feature of the ETV curve that would hint at the presence of the LTTE, and our selection was thus biased towards the contact binaries of periods $<0.6\,$d.

Although the tertiary mass cannot be determined just by the LTTE, the amplitude of the LTTE, calculated by
\begin{equation}
\mathcal{A}_{\text{LTTE}}=\frac{a_{\text{AB}}\sin^{3}i_{2}}{c}\sqrt{1-e_{2}^{2}\cos^{2}\omega_{2}},
\end{equation}
allows the estimation of the mass function, $f(m_{\text{C}})$,
\begin{equation}
f(m_{\text{C}})=\frac{m_{\text{C}}^{3}\sin^{3}i_{2}}{m_{\text{ABC}}^{2}}=\frac{4\pi^{2}a_{\text{AB}}^{3}\sin^{3}i_{2}}{GP_{2}^{2}},
\end{equation}
via the approximation equation given by
\begin{equation}
\mathcal{A}_{\text{LTTE}}\approx 1.1\times10^{-4}f(m_{\text{C}})^{1/3}P_2^{2/3}\sqrt{1-e_{2}^{2}\cos^{2}\omega_{2}},
\end{equation}
where the period and amplitude are in days and the masses are in units of $M_{\sun}$. Assuming the components of the three selected EB candidates all are solar, and thus taking $m_{\text{AB}}$, the total mass of each selected EB candidate, equal to $2\,M_{\sun}$, the calculated minimum masses of their tertiary companions are shown in Table \ref{tab:triple}. 

\section{Discussion and Conclusions}
\label{sec:discuss}
The MOA project was evolved to its second stage in 2006. We presented the first MOA-II catalogue of eclipsing binary (EB) candidates identified in the two MOA Galactic bulge fields, GB9 and GB10, using two seasons of data, from February 2013 to August 2014. The MOA-II EB catalogue contains over 8,000 EB candidates with periods ranging from 0.09 day to 66 days. Most of them are contact or semi-detached binaries of periods $<1$ day. Detached binaries of periods $>30$ days were difficult to identify using just two MOA observational seasons worth of data. The identification rate of long period detached binaries can be increased if the complete datasets are used. However, it will cause substantial computational pressure on the light curve generation and data analysis, especially for the fields with high cadences such as GB9, GB10, and GB5, in which there are no less than 30,000 images in total to date. Another challenge to the EB identification from the \(100\,\text{TB}\) MOA database is to develop a robust automated or machine learning algorithm that can effectively filter out the artifacts and 0.5 day aliases which came up substantially in the manner described in Section~\ref{sec:iden_eb}, and replace the tedious and time-consuming light curve inspection by eye that has limited our study to only two MOA fields from a total of 22 fields. The possible solution to this problem is to develop our own machine learning algorithm, for example, using random decision forests \citep{2011MNRAS.414.2602D} for the MOA database.

Other than contact and semi-detached binaries, there are certain numbers of eccentric binaries, binaries with complicated phase modulations and eclipsing RS CVn type stars as well as a few ultra-short period binaries discovered in the GB9 and GB10 fields. Additionally, three EB light curves were discovered to have extra eclipsing signals under their main eclipsing signals indicating their doubly EB identity. Our interest in this paper also includes searching for triple object candidates in our EB candidates via eclipse time variation (ETV) analysis. The eclipse timing of an EB was performed by the template method in which the eclipse template was constructed using the phenomenological model of EB light curves \citep{2015A&A...584A...8M}. As a result, three contact binaries of periods $<0.6$ day showed ETVs that are well fitted by the light-travel-time effect (LTTE) model. The amplitudes of their LTTE solutions indicate that their tertiary companions are stellar assuming the masses of the inner binaries are $2\,M_{\sun}$. The success in detecting LTTEs in short period EBs in the MOA database of just two observational seasons should verify the value on the further ETV study using the full database in attempt to identify triple systems with longer outer periods.   

\section*{Acknowledgements}
M.C.A. Li acknowledges the contribution of NeSI high-performance computing facilities to the results of this research. NZ's national facilities are provided by the NZ eScience Infrastructure and funded jointly by NeSI's collaborator institutions and through the Ministry of Business, Innovation \& Employment's Research Infrastructure programme. URL https://www.nesi.org.nz. NJR is a Royal Society of New Zealand Rutherford Discovery Fellow. AS is a University of Auckland Doctoral Scholar. TS acknowledges financial support from the Japan Society for the Promotion of Science (JSPS) under grant numbers JSPS23103002, JSPS24253004 and JSPS26247023. NK is supported by Grant-in-Aid for JSPS Fellows. The MOA project is supported by JSPS grants JSPS25103508 and JSPS23340064 and by the Royal Society of New Zealand Marsden Grant MAU1104.





\bibliographystyle{latest}
\bibliography{microlensing,eclipsingBinaries}





\bsp	
\label{lastpage}
\end{document}